\documentclass{aa}  
\usepackage{amsmath}
\usepackage{soul}
\usepackage{tablefootnote}
\usepackage{bm}
\usepackage{xcolor}
\usepackage[normalem]{ulem}
\usepackage[
    range-units=single,         % Formatting ranges with single unit indication: 1 - 2 m
    range-phrase=-,             % Phrase for range: 1 - 2 m vs 1 to 2 m
    separate-uncertainty=true,  % sets +- between value and uncertainty 
    multi-part-units=single     % In expressions with multiple values (multi part numbers) 
] {siunitx}
\DeclareSIUnit\angstrom{Å}
\DeclareSIUnit\gauss{G}
\DeclareSIUnit\Ic{I_c}

\usepackage{graphicx}
\usepackage{txfonts}
\usepackage[colorlinks,linkcolor=blue,citecolor=blue,linktocpage=true,breaklinks, 
plainpages=false,urlcolor=blue]{hyperref}

\definecolor{iimm}{cmyk}{1.0,0.0,1.0,0.3}
\definecolor{ttkk}{cmyk}{0.57,0.3,0.0,0.0}

\usepackage{booktabs} 
\usepackage{multirow} 
\usepackage{cancel} 
\usepackage{adjustbox}
\usepackage{caption} %
\begin{document}

   \title{Multi-height probing of horizontal flows in the solar photosphere}

   \author{T. Kosti\'{c}
          \inst{1}, I. Mili\'{c}\inst{1,2,3}, M. Rempel \inst{4}, B. Welsch \inst{5}, M. Kazachenko \inst{6,7,8}, A. Asensio Ramos \inst{9,10}, B. Tremblay \inst{4,11} 
          }
   \institute{Astronomical Observatory, Volgina 7, 11060 Belgrade, Serbia \and Institute for Solar Physics, Georges Kh\"{o}ler Alee 401a, 79110 Freiburg, Germany \and Department of Astronomy, Faculty of Mathematics, University of Belgrade, Studentski Trg 16-20, 11000, Belgrade, Serbia \and High Altitude Observatory, NSF National Center for Atmospheric Research, Boulder, CO, USA \and University of Wisconsin - Green Bay, Green Bay, WI, USA \and Laboratory for Atmospheric and Space Physics, University of Colorado, Boulder, CO, USA \and Department of Astrophysical and Planetary Sciences, University of Colorado at Boulder, Boulder, CO, USA \and National Solar Observatory, University of Colorado Boulder, Boulder, CO, USA \and Instituto de Astrof\'isica de Canarias (IAC), 
   Avda V\'ia L\'actea S/N, 38200 La Laguna, Tenerife, Spain \and Departamento de Astrof\'isica, Universidad de La Laguna, 38205 La Laguna, Tenerife, Spain \and Environment and Climate Change Canada, Science and Technology Branch, Meteorological Research Division, Dorval, Québec, Canada\\
   \email{teodor@aob.rs}}
   
   \date{Received ; accepted }

% \abstract{}{}{}{}{} 
% 5 {} token are mandatory
 
  \abstract
   {Optical flow methods aim to infer horizontal (transverse, in the general case) velocities in the solar atmosphere from the temporal changes in maps of physical quantities, such as intensity or magnetic field. So far, these methods have mostly been tested and applied to the continuum intensity and line-of-sight (LOS) magnetic field in the low to mid-photosphere.}
   {We tested whether simultaneous spectropolarimetric imaging in two magnetically sensitive optical spectral lines, which probe two different layers of the solar atmosphere (the photosphere and the temperature minimum), can help constrain the depth variation of horizontal flows.}
   {We first tested the feasibility of our method using Fourier local correlation tracking (FLCT) to track physical quantities at different optical depths ($\log\tau_{500}={-1,-2,-3,-4}$) in an atmosphere simulated with the MURaM code. We then inferred the horizontal distribution of the LOS magnetic field component from synthetic spectropolarimetric observations of Fe\,I\,525.0\,nm and Mg\,I\,b2 spectral lines, applied FLCT to the time sequence of these synthetic magnetograms, and compared our findings with the original height-dependent horizontal velocities. }
   {Tracking the LOS magnetic field component (which coincides with the vertical component at the disk center) yields horizontal velocities that, after appropriate temporal and spatial averaging, agree excellently with the horizontal component of the simulated velocities, both calculated at constant $\tau_{500}$ surfaces, up to the temperature minimum ($\log\tau_{500}=-3$). When tracking the temperature at constant $\tau_{500}$ surfaces, this agreement already breaks down completely at the mid photosphere ($\log\tau_{500}=-2$). Tracking the vertical component of the magnetic field inferred from synthetic observations of the Fe\,I\,525.0\,nm and the Mg\,I\,b2 spectral lines yields a satisfactory inference of the horizontal velocities in the mid-photosphere ($\log\tau_{500}\approx-1$) and the temperature minimum ($\log\tau_{500}\approx-3$), respectively.}
   {Our results indicate that high-spatial-resolution spectropolarimetric imaging in solar spectral lines can provide meaningful information about the horizontal plasma velocities over a range of heights.} 
     \titlerunning{Multi-height flow tracking}
    \authorrunning{Kosti\'{c} et al.}

   \maketitle

%________________________________________________________________

\section{Introduction}
\label{sec:intro}
 
The structure and dynamics of the solar lower atmosphere are driven by interaction between plasma motions and the magnetic field on various spatial scales. In the quiet Sun (QS), in weakly magnetized regions where the magnetic field strength is often below the equipartition value between the magnetic and dynamic pressure (i.e., well below  $\SI{1}{kG}$), the magnetic field structure is largely determined by plasma motions. By probing plasma dynamics, we can understand the transport and amplification of the magnetic field, as described by the induction equation for an ideal, infinitely conducting fluid: 
\begin{equation}
\label{eq:induction}
    \frac{d\vec{B}}{dt} = \nabla \times (\vec{v} \times \vec{B}),
\end{equation}
where $\vec{v}$ is the velocity vector and $\vec{B}$ is the magnetic field vector. Therefore, the velocity field is a crucial parameter for understanding solar magnetism on various scales, including, but not limited to, mean-field theories of the solar dynamo \citep[e.g.,][]{Pauldynamo2010}, small-scale dynamo processes \citep[e.g.,][]{Rempel2014}, energy transport from the photosphere upward \citep{Welsch_2015_plage_poyinting, Tilipman2023}, and magnetohydrodynamic (MHD) wave studies \citep{Skirvin_2024ApJ_MHDwaves}. Estimating plasma velocities also allows us to quantify the energy and mass transport associated with various scales of solar convection \citep[e.g.,][]{Nordlund_lrsp_2009}, waves \citep[e.g.,][]{arregui2015}, or swirls \citep[][]{swirl2019}.

The line-of-sight (LOS) component of the plasma velocity is typically estimated from the Doppler effect by using simple line core fitting, bisector techniques, or spectropolarimetric inversion techniques, which provide depth-dependent variations of velocity and other physical parameters \citep[e.g.,][]{Michiel_Jaime_2017}. The transverse velocity component (corresponding to horizontal motions at the disk center) is often probed through optical flows.
The term optical flow, originally coined for computer vision applications \citep[see the discussion in][]{Schuck_2006_DAVE}, refers to the apparent motion of a feature, inferred from images taken at different times \citep{Chae2008}. Hence, tracking optical flows involves using one or more techniques to recover the transverse velocity field of the plasma. To derive horizontal velocities using tracking methods, a time series of continuum intensity images or LOS magnetograms (maps of the magnetic field $B_{\rm LOS}$ at different times) is required \citep{Welsch2007, Verma2013}. The choice between the two is often based on where the tracking is performed (e.g., quiet Sun versus active regions). 
Over the years, various methods for tracking flows have been developed, but the most commonly used and well-known optical flow method is local correlation tracking \citep[LCT;][]{NovemberSimon1988}. As the name suggests, it is a local method: the velocity at a given pixel is based on image values within a small area centered on that pixel \citep{Chae2008}. 

Fourier local correlation tracking \citep[FLCT;][]{Welsch2004, Welsch2008}, that is, the use of Fourier techniques to calculate correlations, is a faster and more accurate variation of LCT that has been widely applied in solar atmosphere diagnostics. For example, \cite{Welsch2004} applied FLCT to measure the rate of magnetic helicity injected into the corona through the photosphere. \cite{swirl2019} applied FLCT to intensity images obtained by the Solar Optical Telescope (SOT) aboard the Hinode satellite to infer the average radius and rotation speed of photospheric and chromospheric swirls. Moreover, \cite{Li_2021} used FLCT to estimate the velocities of supra-arcade downflows (SADs), the plasma voids associated with flare loops. \cite{Fisher_2020} used the FLCT code to determine horizontal velocities and their contribution to the noninductive electric field as part of their PTD-Doppler-FLCT-Ideal (where PTD stands for poloidal-toroidal decomposition) software library, $\mathrm{PDFI\_SS}$, which computes the electric field at the Sun's photosphere \citep{Kazachenko2014,Lumme2019}. 

Multiple studies have benchmarked the performance of the LCT and FLCT techniques by applying them to synthetic intensity maps and magnetograms produced by various magnetohydrodynamic (MHD) simulations of the solar photosphere \citep[see, e.g.,][]{Rieutord2001, Verma2013,Kazachenko2014, bendza2018,Afanasyev2021}. These benchmarks aim to synthesize observables, apply tracking techniques to them, and compare the results with the original horizontal velocities. For example, \citet{Loptien2016} explained some of the biases in LCT techniques by applying FLCT to continuum images obtained with the Helioseismic and Magnetic Imager (HMI) on board the Solar Dynamics Observatory \citep[SDO;][]{Schou2012} and to images generated from STAGGER code simulations \citep{Stein2012}, showing that both exhibit a shrinking-Sun effect of comparable magnitude. The term ``Shrinking-Sun'' refers to a systematic error that appears as a flow converging towards the disk center, which is superimposed on real flows and can reach up to 1\,km/s. In general, these benchmarks focus on applying FLCT either to continuum images or to photospheric magnetograms.

In this work, we test the feasibility of recovering plasma flows at multiple atmospheric heights, from the base of the photosphere to the temperature minimum. To this end, we applied FLCT to the time series of the simulated continuum intensity, temperature, $T$, and the vertical component of the magnetic field, $B_z$ at multiple optical depths, from $\log\tau_{500}=0$ to $\log\tau_{500}=-4$, and compared the resulting velocities with the original simulation values. We then repeated the same test by tracking magnetograms inferred from synthetic observations of the neutral iron line at 525.02\,nm and the b2 line of neutral magnesium at 517.2\,nm, which probe the mid-photosphere ($\log\tau_{500}\approx-1$) and the temperature minimum ($\log\tau_{500}\approx-3$), respectively. The motivation for using spectral lines lies in their sensitivity to different depths in the solar atmosphere. Specifically, the monochromatic absorption coefficient of the solar plasma increases rapidly with wavelength toward the line center, with substantial changes occurring on picometer scales. This leads to different formation heights for radiation observed at the continuum and line wavelengths. Therefore, interpreting time-dependent multi-line observations using spectropolarimetric inversions \citep{Michiel_Jaime_2017} should allow tracking of flows across different layers of the solar atmosphere (different values of optical depth $\tau_{500}$, i.e, height $z$). Spectral lines also provide a means to infer the LOS velocity through the Doppler effect, and thus probe all three components of the velocity vector. This is especially important in the context of existing and upcoming high-resolution observations of the solar atmosphere obtained with imaging spectropolarimeters such as SST/CRISP \citep{CRISP2008}, SUNRISE/TuMag \citep{TuMag}, and DKIST/VTF \citep{vtf}, or with integral field units such as Microlensed Hyperspectral Imager \citep[MiHI;][]{vanNoort2022MiHI} or Helium Spectropolarimeter \citep[HeSP, see][]{Leenaarts2025hesp}. 
Both the creation and interpretation of spectropolarimetric data, and the analysis of multi-height flows, distinguish this work from previous studies that used simulated data to characterize the fidelity of flow reconstruction techniques \citep[e.g.,][]{Verma2013,bendza2018}.

This paper is organized as follows.
In Sect.\,\ref{sec:methods}, we describe the basics of the FLCT method, the simulations we used, and the choice of quantities to track. Section\,\ref{sec:results} presents the verification of the FLCT applied to continuum intensity and the vertical component of the magnetic field at the base of the photosphere for these specific simulations, followed by the results of tracking applied to the time series of temperature and $B_z$ maps at different optical depths ($\log \tau_{500} = -1$ to $\log\tau_{500} = -4$). Section \ref{sec:tumag} presents a proof-of-concept study performed on synthetic SUNRISE/TuMag data in two spectral lines. Finally, Sect. \ref{sec:conclusions} summarizes our conclusions and outlines future tests and applications.

\section{Methods}
\label{sec:methods}
\subsection{FLCT}
\label{ssec:flct}

Fourier local correlation tracking (FLCT) is a variation of the often used and well-known optical flow method in the solar research community, LCT \citep{Welsch2004, Welsch2008}. The FLCT code uses Fourier cross-correlation of the quantity $X_i(x,y,t_i)$ with the quantity $X_f(x,y,t_f)$ (where the subscripts $i$ and $f$ refer to the initial and final time steps, respectively) near a given centered pixel to determine the two-component displacement $\Delta\vec{s}$ of the structure near that pixel. Division by $\Delta t$ results in a velocity estimate, $\vec{v}$. The correlation is localized by weighting each image with a Gaussian window function,
\begin{equation}
    F(r) = \exp(-r^2/\sigma^2).
    \label{apodization}
\end{equation}
Here, $r$ denotes the distance from the pixel at which the vector $\Vec{v}$ is estimated \citep{Welsch2012}. The full width at half maximum (FWHM) of the Gaussian weighting is equal to $1.665\sigma$, and FWHM is used instead of $\sigma$ throughout this paper. In other words, FLCT applies an apodization or windowing function, with a user-defined FWHM, to the initial and final images (maps) of the time series and determines the most likely displacement for each pixel from the correlations. The size of the apodizing window (i.e., the Gaussian-shaped weighting function) should approximately match the spatial scale of the structure being tracked.

It is important to note that optical flows do not always represent actual horizontal plasma motions. For example, as discussed by \cite{Demoulin_Berger_2003}, optical tracking methods applied to magnetograms may incorporate apparent motions that do not correspond to true horizontal plasma motions, but instead arise from the emergence of inclined magnetic structures. This occurs because the methods in question consider only the vertical component of the magnetic field and assume that its changes are driven by horizontal plasma motion. However, the vertical plasma velocity can also contribute because when a tilted flux tube rises through the photosphere, its intersection with the photosphere moves. The authors concluded that tracking methods that follow photospheric footpoints of flux tubes measure a weighted sum of both vertical and horizontal velocities, rather than only horizontal \citep[see also][]{Welsch2004, Welsch2012}. Nevertheless, optical flows are found to be highly correlated with plasma velocity fields at scales larger than \SI{2.5}{Mm}, and the correlation can be further improved by computing time averages of the inferred instantaneous velocities \citep{Rieutord2001}. 

The size of the apodizing window is crucial for the tracking outcome. Firstly, it determines the spatial scales to be tracked and the fineness of flow structures that can be recovered. Secondly, it strongly affects the computing time of the FLCT code. \cite{Chae2008} proposed selecting a smaller apodizing window to obtain more detailed velocity maps and reduce computational demands. If high-cadence images are available, smaller sampling windows track fast-moving, fine structures but cannot measure the horizontal proper motions of larger, coherent structures. At the other limit, larger sampling windows enable plasma motion tracking on larger scales. It should be noted that FLCT typically underestimates flow speeds and that this underestimation increases as the FWHM of the apodizing window decreases \citep[e.g.,][and also confirmed by our results]{Verma2013}.

\subsection{Simulations}
\label{ssec::sims}

We used a small-scale dynamo simulation of the solar photosphere performed with the Max-Planck-Institute for Aeronomy/University of Chicago radiation magnetohydrodynamics (MURaM) code \citep{MURaM2005,Rempel2014}. Specifically, we restarted the simulation from case O16bM in \citet{Rempel2014}, which used a bottom boundary condition (symmetric in B) that accounts for magnetic field recirculation from the deeper convection zone. This setup was shown to be consistent with quiet Sun observations presented in \citet{danilovic2016}. The simulation domain was extended upward by approximately \SI{500}{\kilo\m}, resulting in a spatial domain of $24.576 \ \times \ 24.576 \ \times \ 8.192 \ \SI{}{\mega\m^3}$. The upper boundary was located approximately $\SI{2}{\mega\m}$ above the average layer $\tau_{500} = 1$. The grid spacing was $\SI{16}{km}$ in all three dimensions, corresponding to a horizontal extent of $1536\times1536$ pixels. The simulation used a non-gray radiative transfer treatment with four opacity bins \citep[e.g.][]{Nordlund_1982_simulations} to allow a more accurate synthesis of photospheric spectral lines. The simulation covered one hour of solar time, with 2D maps of physical parameters output at a cadence of 10 seconds and 3D cubes of physical parameters output at a cadence of 30 seconds.

\begin{figure*}
    \centering
    \includegraphics[width=\textwidth]{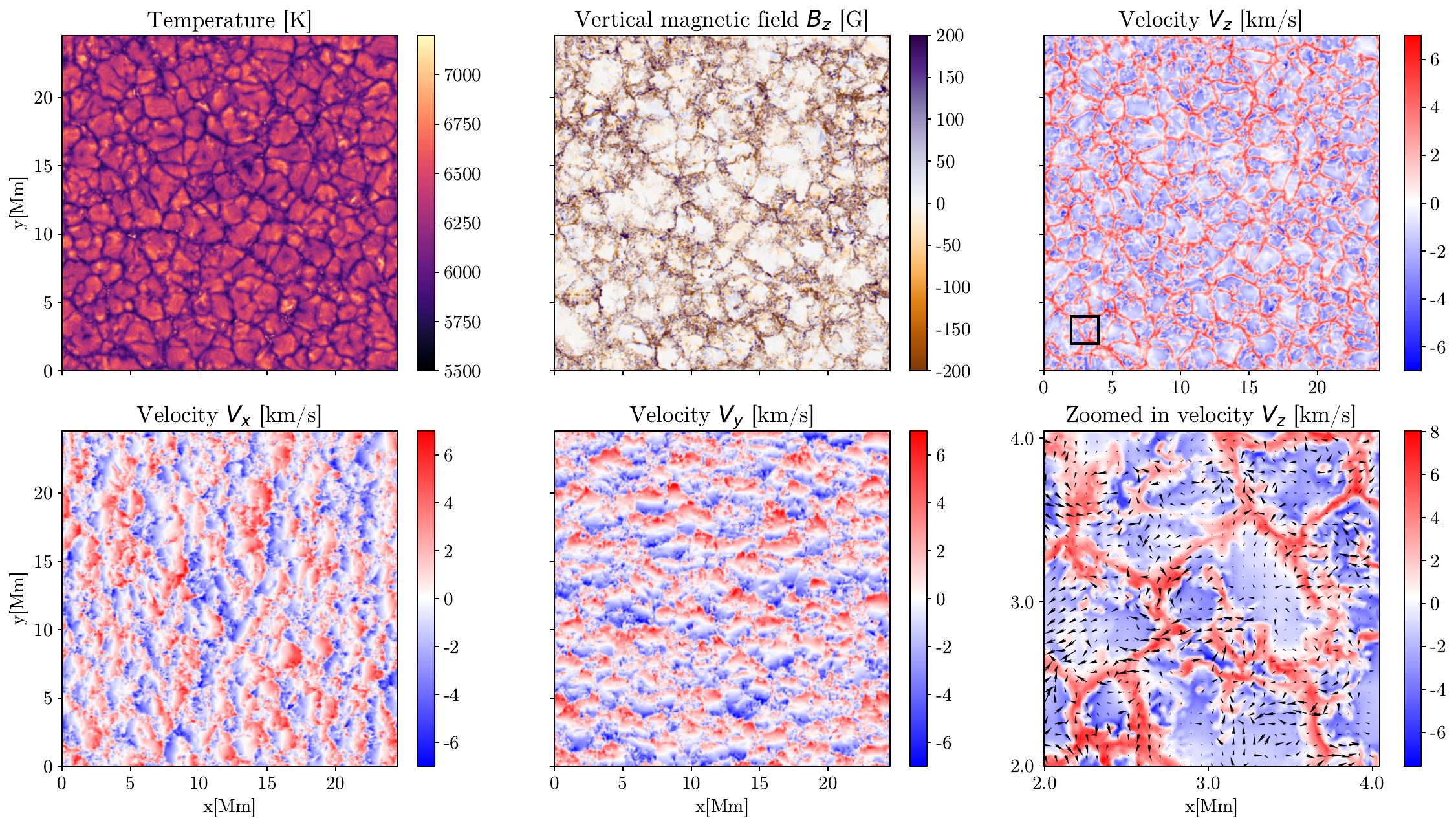}
    \caption{Temperature, vertical magnetic field, and the three components of the velocity at $\log \tau_{500} = 0$ for the first considered timestep of the simulation. The black square in the upper-right panel marks the region shown in the lower-right panel. Arrows in the lower-right panel indicate horizontal velocities.}
    \label{fig:simlogtau0}
\end{figure*}

For spectrum synthesis, we used 31 3D cubes (snapshots) containing the spatial variations of physical parameters (temperature, $T$, total gas pressure, $p_g$, velocity vector, $\vec{v}$, and magnetic field vector, $\vec{B}$) output every 30 \,s of solar time, equivalent to a total of 15 solar minutes. These cubes are defined on the same geometrical mesh as that used to solve the MHD equations. Additionally, we used the emergent intensity, calculated with MURaM, in the direction $\theta=0$ at wavelength $\lambda = \SI{500}{nm}$, and the values of the velocity, temperature, and magnetic field vectors at iso-optical-depth surfaces corresponding to $\log \tau_{500} = \{0, -1, -2, -3, -4\}$. For this observing direction, the LOS magnetic fields coincide with the $B_z$ component. The optical depth is defined as
\begin{equation}
    \tau_{500}(z) = \int_z^{\infty} \chi(z) \, dz~,
    \label{tau}
\end{equation}
where $\chi(z)$ is the opacity at a reference wavelength, in this case, $\SI{500}{nm}$. These layers span heights from the base of the photosphere up to the temperature minimum. The spatial distribution of temperature, velocity, and the vertical magnetic field at $\log \tau_{500} = 0$ for the first snapshot of the series is shown in Fig.\,\ref{fig:simlogtau0}.

\subsection{Synthetic spectra and degradation}
\label{ssec:synt}

To create synthetic SUNRISE/TuMag observations, we calculated the emergent intensity of each pixel in the direction $\theta = 0$ (disk center) for a range of wavelengths containing the line of neutral iron at 525.02 nm and the line of neutral magnesium at 517.2 nm (hereafter Fe\,I\,$\SI{525.0}{nm}$ and Mg\,I\,b2). The Fe I line is a photospheric spectral line with a large Land\'{e} factor ($g=3$), which makes it a prime candidate for studies of photospheric magnetism. The Mg\,I line is a strong line with moderate sensitivity to the magnetic field ($g=1.75$), which probes the photosphere and the temperature minimum. The choice of these lines was motivated by the observations performed with the TuMag instrument \citep{TuMag} at the SUNRISE \citep{Sunrise:2010, SUNRISE} balloon-borne telescope, which completed its third successful flight in July 2024 \citep{KorpiLagg2025sunriseIII}.   

For synthesis, we used spectropolarimetric non-local thermodynamic equilibrium (NLTE) analytically powered inversion \citep[SNAPI;][]{Milic2018snapi}, a 1D radiative transfer and inversion code that synthesizes a large number of spectra using message passing interface (MPI) parallelization. The code accounts for continuum and spectral line opacity, polarization due to the Zeeman effect, and NLTE effects in the Mg\,I\,b2 line. It accurately reproduces the observed shape of the Mg\,I\,b2 line \citep{Vukadinovic2022mg}. Neutral iron is overionized in the solar photosphere due to UV radiation, and rigorous modeling would require an NLTE approach \citep{Smitha_nlteI}. Such a calculation would be prohibitively expensive for the entire time series. Furthermore, because our study is primarily methodological,  
highly realistic, and precise modeling is beyond the scope of this paper. 

\begin{figure*}
    \centering
    \includegraphics[width=0.99\textwidth]{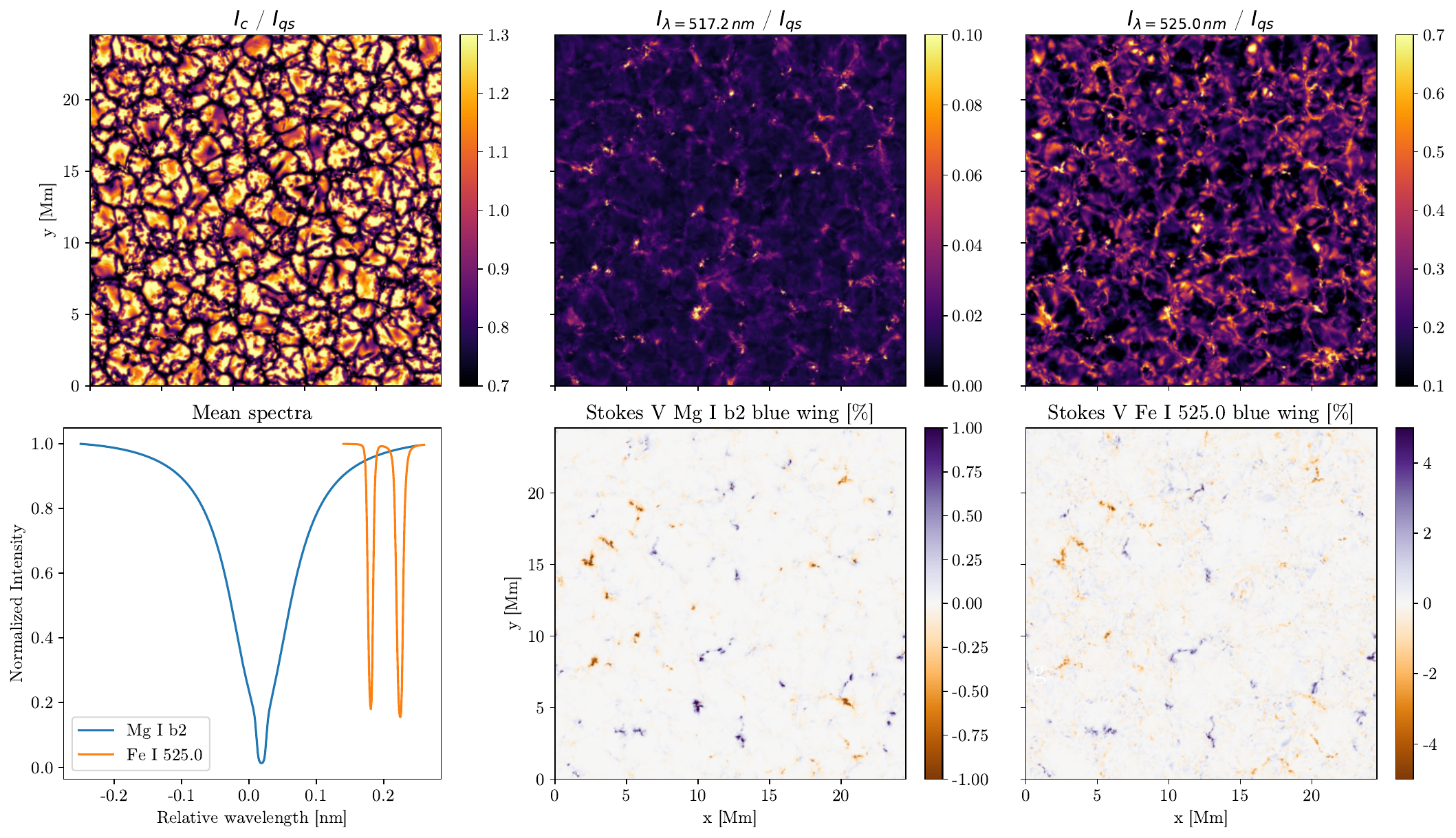}
    \caption{Synthetic observables calculated for the first snapshot in the series. Top left: Continuum image at 517.2\,nm. Top middle: Nominal line core of the Mg\,I\,b2 line. Top right: Nominal line core of Fe\,I\,525.0\,nm line. Bottom left: Spatially averaged spectra of the considered spectral lines, Fe\,I lines are shifted for clarity. Bottom middle: Circular polarization in the wing of the Mg\,I line. Bottom right: Circular polarization in the wing of the Fe\,I line. The upper-row panels are shown in units of the mean quiet-Sun continuum.}
    \label{fig:multi_lambda_line}
\end{figure*}

The spectra were synthesized using every second pixel along $x$ and $y$,  increasing the spatial step to $32$\,km and making it comparable to the TuMag pixel scale \citep[28\,km;][]{TuMag}. The spectra of the two lines were synthesized with 1\,pm sampling, using 121 wavelengths for the Fe\,I\,525.0\,nm line and 501 wavelengths for the Mg\,I\,b2 line. The spectra were synthesized for all 31 time steps. The total number of resulting polarized spectra is therefore $768\times 768 \times 31 \approx 18 \times 10^6$. We synthesized two neutral iron lines around 525.0\,nm, but only considered the more magnetically sensitive one, $3d^64s^2 \rightarrow 3d^6(5D)4s^4p(3P^0)$, with the central wavelength of 525.0208\,nm. 
Figure\,\ref{fig:multi_lambda_line} shows the synthetic intensity in the continuum, nominal cores of the two lines, and the circular polarization in the wings of the two lines. The map of the circular polarization in the Mg\,I\,b2 line was obtained by averaging Stokes $V$ over a 50\,${\mathrm m\AA}$ interval, centered at the wavelength point 70\,${\mathrm m\AA}$ to the blue of the nominal line core. The map of the circular polarization in the Fe\,I\,525.0\,nm line was obtained similarly, but centered at the wavelength point 50\,${\mathrm m\AA}$ to the blue of the nominal line core. Section\,\ref{sec:tumag} presents the methods used to extract relevant physical quantities from these synthetic observations.

\section{Results}
\label{sec:results}
\subsection{Tracking continuum intensity and magnetic field}
\label{ssec:track_og}

To test the applicability of the FLCT technique for inferring horizontal velocities, we first applied it to the continuum intensity maps at $\lambda = 500$\,nm, which are the output of the simulation, and to magnetograms, i.e., maps of the vertical component of the magnetic field vector at constant-$\tau_{500}$ corrugated surfaces, calculated using the MURaM code. We applied FLCT to the entire one-hour time series, which yielded a map of horizontal velocities for each pair of consecutive intensitygrams and magnetograms. We then compared the obtained velocities with horizontal velocities from the simulation (calculated as the average between two consecutive snapshots) at constant-$\tau_{500}$ surfaces. Direct comparison of these quantities yields relatively poor results \citep[e.g.][]{Verma2013, bendza2018}. This occurs because the apodizing window used in FLCT automatically imposes a minimal spatial scale at which we can probe horizontal flows. Therefore, to compare the inferred and original velocities, the latter need to be convolved with a Gaussian filter that corresponds to the apodizing window used by FLCT. As shown by \citet{Verma2013} and \citet{bendza2018}, the best agreement between the inferred and original velocities is achieved when both are averaged over a time interval that is much longer than the cadence of the analyzed images. For this dataset, we find that the best agreement occurs when both the inferred and original velocities are averaged over 15 minutes. Thus, each 15-minute interval yielded a single map of horizontal velocities, which was then compared to a single map of spatially convolved and temporally averaged simulated velocities. All comparisons in this and the following subsections are based on snapshots from the first 15 minutes of the simulated atmosphere.  For the analysis of the synthetic spectra, we focused on the same 15-minute interval, sampled at a cadence of 30s cadence, corresponding to 31 cubes in total. 
For comparison, \citealt{Verma2013} averaged over 15, 30, 60, 90, and 120-minute intervals.
This time averaging step implies that, when applying FLCT, a compromise must be made between spatial and temporal resolution. This is a known limitation of FLCT techniques, which can be mitigated using recent machine learning-based tracking approaches \citep{aar_deepvel}. Hereafter, we refer to the temporally averaged velocity inferred by FLCT as $\vec{v}_f = (v_x, v_y)_f$, with the subindex $f$ indicating that $\vec{v}_f$ was obtained using FLCT from images of the quantity in parentheses. The temporally averaged and spatially filtered velocity from the simulation is denoted as $\vec{v}_o = (v_x, v_y)_o$.

\begin{figure}[h!]
    \centering
    \includegraphics[width=0.49\textwidth]{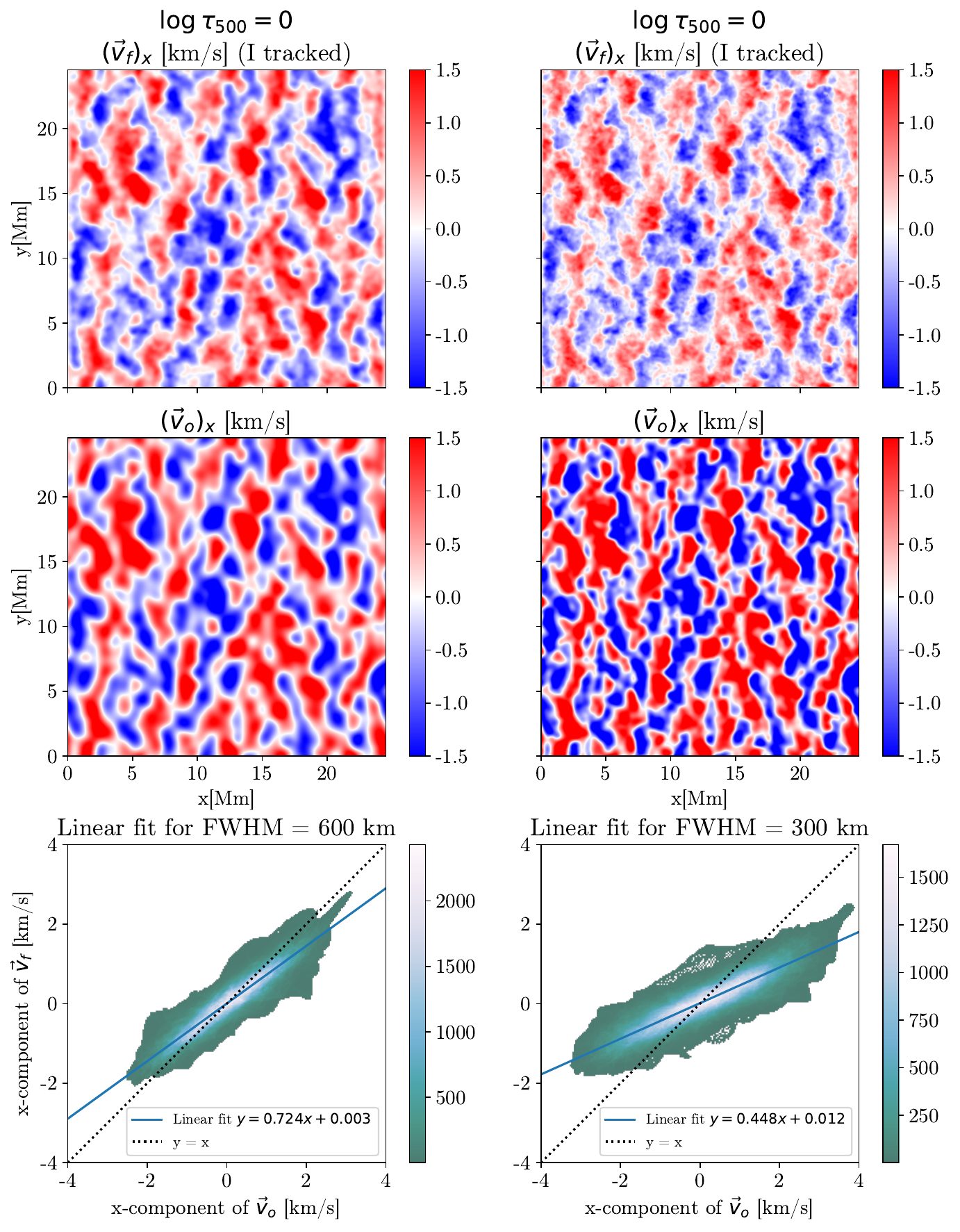}
    \caption{Comparison between the $x$ component of the horizontal velocity retrieved by FLCT applied to 
    intensity maps and the simulated velocity at $\log\tau_{500}=0$. Color maps indicate the magnitude of velocity. Bottom row: Scatter plot showing the least-squares fit between $\vec{v}_o$ and $\vec{v}_f$. Results are shown for two different sizes of apodizing windows: left, $\SI{600}{km}$; right, $\SI{300}{km}$.}
    \label{fig:track_I}
\end{figure}

\begin{figure}[h!]
    \centering
    \includegraphics[width=0.49\textwidth]{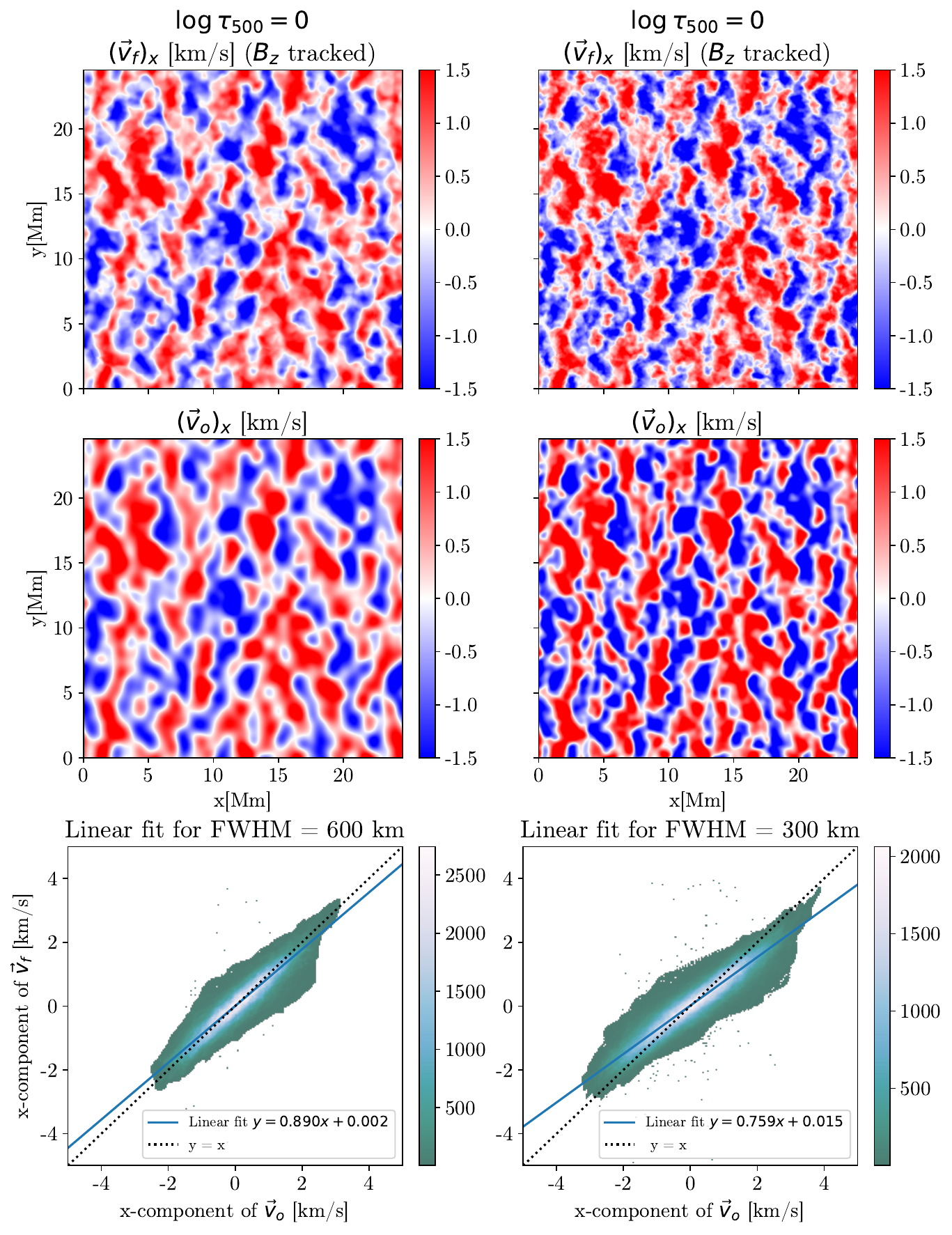}
    \caption{Same as Fig.\,\ref{fig:track_I}, but for FLCT applied to $B_z$ at $\log\tau_{500}=0$.}
    \label{fig:track_B}
\end{figure}

Figure\,\ref{fig:track_I} shows the comparison between the $x$ component of $v_f$, inferred by tracking the continuum intensity, and the $x$ component of $v_o$, for two different sizes of the apodizing windows: $\SI{600}{km}$ and $\SI{300}{km}$. Figure\,\ref{fig:track_B} presents the same comparison for FLCT applied to $B_z$ maps at $\log\tau_{500}=0$ (i.e., the base of the photosphere). Because our motivation is the interpretation of high-resolution observations  \citep[while][have tested apodizing windows in range 400-1800 km with a step of 200 km in order to benchmark the limitations of FLCT]{Verma2013}, we did not test apodizing windows larger than $\SI{600}{km}$.
Apodizing windows significantly below $\SI{300}{km}$ did not yield satisfactory agreement with the simulated velocities. This occurs because the assumptions of FLCT and the time averaging used in the comparison procedure do not resolve plasma motion on scales significantly below the spatial scales of the granules \citep[as reported by][for atmospheres simulated using CO5BOLD]{Verma2013}.

\begin{table}[h!]
\caption{Pearson's correlation coefficients between simulated velocities and FLCT-derived velocities at the base of the photosphere.}

\begin{center}
\begin{adjustbox}{width=0.48\textwidth}
  \begin{tabular}{ @{} lcSSSSSS @{} }
    \toprule
    \multirow{2}{*}{} & 
    \multirow{2}{*}{FWHM[km]} & 
      \multicolumn{2}{c}{O} &
      \multicolumn{2}{c}{OT} &
      \multicolumn{2}{c}{OTC} \\
      && {$r_{v_x}$} & {$r_{v_y}$} & {$r_{v_x}$} & {$r_{v_y}$} & {$r_{v_x}$} & {$r_{v_y}$} \\
      \midrule
    \multirow{2}{*}{Intensity} & 600 & 0.56 & 0.55 & 0.82 & 0.81 & 0.94 & 0.93 \\
     & 300 & 0.51 & 0.51 & 0.83 & 0.83 & 0.92 & 0.91 \\
      \cmidrule(l){2-8} 
    \multirow{2}{*}{$B_z$} & 600 & 0.61 & 0.58 & 0.83 & 0.83 & 0.94 & 0.94 \\
    & 300 & 0.68 & 0.64 & 0.90 & 0.90 & 0.93 & 0.93 \\
    \bottomrule
  \end{tabular}
  
    \label{table:1}
    \end{adjustbox}
  \end{center}
  \tablefoot{The FLCT velocities are derived from the simulated intensity or the magnetic field at $\log\tau_{500} = 0$. O: raw data; OT: velocities averaged over a 15-minute interval; OTC: velocities averaged over a 15-minute interval and convolved with Gaussian filter corresponding to the apodizing window used for FLCT.}
\end{table}

Similar to previous studies, we find that, after spatial convolution and temporal averaging, the inferred and original velocities show excellent correlation (exceeding 0.9 for both $x$ and $y$ components and for both apodizing-window sizes, whether derived from intensity or $B_z$ (see Table\,\ref{table:1}). However, FLCT underestimates the plasma velocity. For the apodizing window of $\SI{600}{km}$, the inferred velocity is approximately three-fourths of the original value (see the bottom-left panels of Fig.\,\ref{fig:track_I}), while for the $\SI{300}{km}$ window, it decreases to two-fifths (see the bottom-right panels of Fig.\,\ref{fig:track_I}). Although this effect has been reported previously \citep[see][]{Verma2013}, its origin is not fully understood. End-to-end studies of this kind can help calibrate this relationship and enable a more accurate interpretation of FCLT results applied to real-life observations. 

The continuum intensity carries information from the iso-optical-depth surface at $\log \tau_{500}=0$, corresponding to the base of the photosphere. 
To track the horizontal velocity at higher layers of the atmosphere, we need to identify the observables that probe these layers. In principle, spectroscopic or spectropolarimetric imaging \citep{Iglesias2019review} in strong spectral lines enables the inference of physical parameters, such as temperature and magnetic field, at multiple atmospheric layers using spectropolarimetric inversion techniques \citep{Iniesta2016rev, Michiel_Jaime_2017}. An important caveat is that spectroscopic and spectropolarimetric observations carry depth information on the optical depth scale. Thus, horizontal information inferred at a given optical depth corresponds to a corrugated surface in geometrical space. Accurate conversion from optical depth to geometrical height therefore requires additional physical constraints \citep[e.g.,][]{borrero_mhs_I, borrero_mhs_IV}.

\begin{figure}
    \centering
    \includegraphics[width=0.495\textwidth]{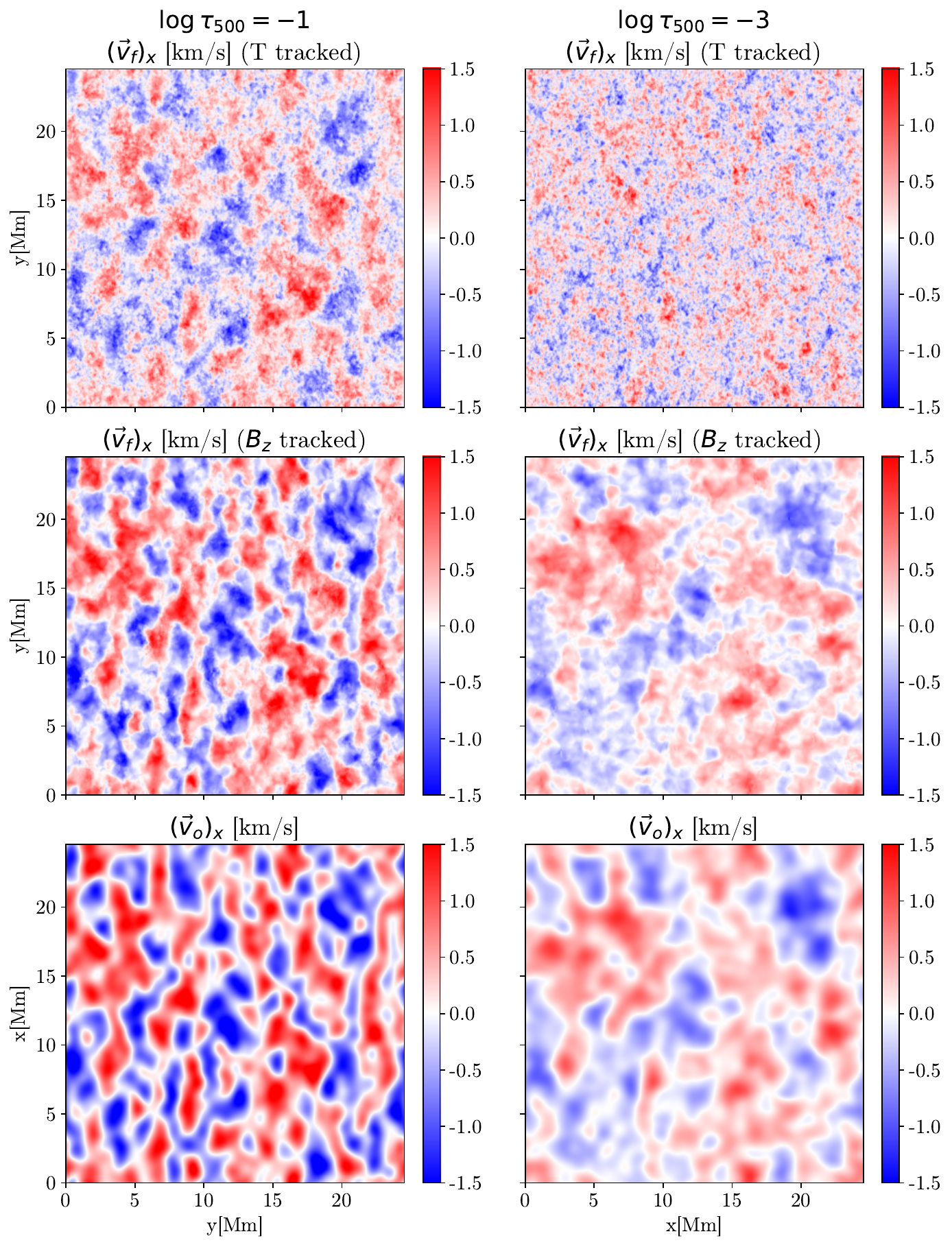}
    \caption{Comparison between the $x$ component of the horizontal velocity retrieved by FLCT using FWHM = 600\,km, applied to temperature (top panel), and $B_z$ (middle panel), and the simulated velocity (bottom panel) at $\log\tau_{500}=\{-1, -3\}$.} 
    \label{fig:track_BzTemp}
\end{figure}

\begin{table}[h!]
\caption{Pearson's correlation coefficients between simulated velocities and FLCT velocities at the mid-photosphere and temperature minimum.}
\begin{center}
\begin{adjustbox}{width=0.48\textwidth}
  \begin{tabular}{ @{} lcSSSSSSSS @{} }
    \toprule
    \multirow{2}{*}{} &
    \multirow{2}{*}{FWHM[km]} &
      \multicolumn{2}{c}{$\log\tau_{500} = -1$} &
      \multicolumn{2}{c}{$\log\tau_{500} = -2$} &
      \multicolumn{2}{c}{$\log\tau_{500} = -3$} &
      \multicolumn{2}{c}{$\log\tau_{500} = -4$} \\
      && {$r_{v_x}$} & {$r_{v_y}$} & {$r_{v_x}$} & {$r_{v_y}$} & {$r_{v_x}$} & {$r_{v_y}$} & {$r_{v_x}$} & {$r_{v_y}$}\\
      \midrule
     \multirow{2}{*}{T} & 600 & 0.72 & 0.73 & 0.38 & 0.41 & -0.05 & -0.06 & 0.03 & 0.04\\
     & 300 & 0.64 & 0.65 & 0.33 & 0.35 & 0.05 & 0.05 & 0.11 & 0.12\\
      \cmidrule(l){2-10} 
    \multirow{2}{*}{$B_z$} & 600 & 0.90 & 0.90 & 0.87 & 0.88 & 0.86 & 0.85 & 0.87 & 0.88\\
    & 300 & 0.91 & 0.91 & 0.90 & 0.90 & 0.86 & 0.85 & 0.85 & 0.86\\
    \bottomrule
  \end{tabular}
  
    \label{table:2}
\end{adjustbox}
  \end{center}
  \tablefoot{Here, FLCT velocities are recovered from either temperature or $B_z$,  at $\log\tau_{500}=\{-1, -2, -3, -4\}$. Correlations are shown only for filtered simulation velocities.}
\end{table}

\subsection{Tracking temperature and magnetic field at multiple optical depths}
\label{ssec::track_multi}

To verify whether spectral line observation can probe horizontal velocities at multiple heights, we applied FLCT to the temperature and $B_z$ at four layers above the base of the photosphere, $\log{\tau_{500}} = \{-1, -2, -3, -4\}$, taken directly from the simulation. We then compared the results with the corresponding velocities in the simulation. Temperature was chosen as the tracking parameter because it is the closest possible quantity to intensity in a spectral line, formed at a specific depth, while $B_z$ was chosen because polarization in different spectral lines is sensitive to the magnetic field at different depths. Tracking these two parameters provides a reference for what can be expected when tracking multi-depth physical quantities inferred from observations.

\begin{figure}
    \centering
    \includegraphics[width=0.495\textwidth]{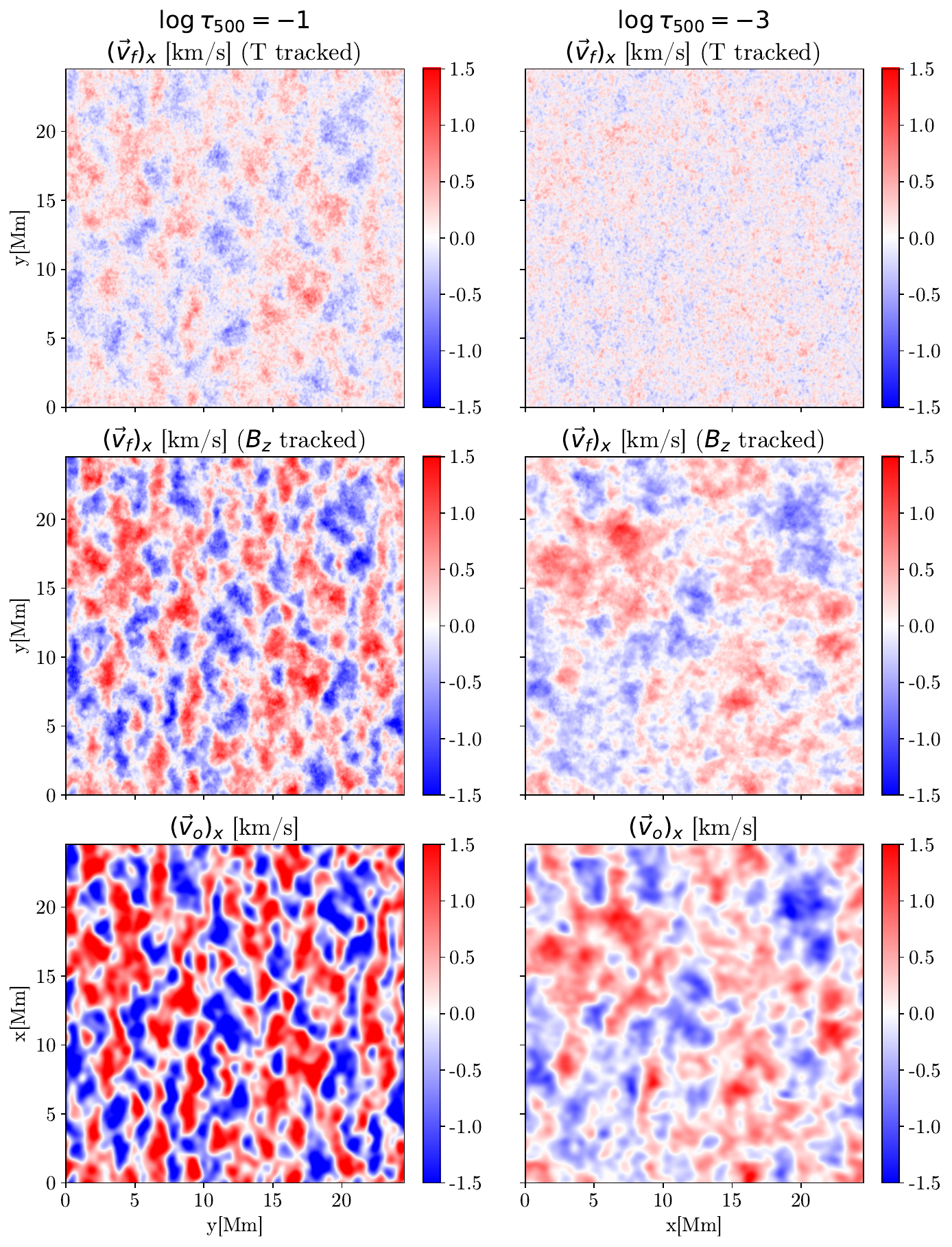}
    \caption{Same as Fig.\,\ref{fig:track_BzTemp}, except that FWHM = 300 km.} 
    \label{fig:track_BzTemp300}
\end{figure}

Figures\,\ref{fig:track_BzTemp} and \ref{fig:track_BzTemp300} compare $v_f$, inferred by tracking temperature and $B_z$, with $v_o$ at $\log\tau_{500}=-1$ and $\log\tau_{500}=-3$, for apodizing windows of 600 and 300\,km. We chose a cadence of 30\,s to match that of synthetic spectropolarimetric observations. Table\,\ref{table:2} lists the Pearson correlation coefficients for all four optical depths and both velocity components. For temperature, the correlation clearly decreases with height. It drops from  $r \approx 0.72$ at $\log\tau_{500} = -1$ to $r \approx 0$ at $\log\tau_{500} = -3$ and higher. By contrast, $B_z$ is a suitable tracking parameter, as FLCT-inferred velocities exhibit a relatively high correlation with the original velocities at all optical depths analyzed. Closer inspection confirms that, structurally, the FLCT velocities derived from $B_z$ closely  match the original velocities, unlike in the case of temperature. The highest layer considered, at $\log \tau_{500} = -4$, may be affected by the choice of the upper boundary of the computational box. Comparison with similar MURaM simulations, which include much higher atmospheric layers (chromosphere and corona), reveal only small differences at $\log\tau_{500} = -4$, prompting us to retain this layer in our analysis.

The stark difference between horizontal velocities obtained by tracking $T$ and $B_z$ arises because temperature is related to internal energy and experiences strong contributions from radiation transport, making its behavior far from advective. For $B_z$, we suppose that the changes are largely due to advection governed by the induction equation \eqref{eq:induction}, providing more robust features for tracking. Therefore, we conclude that FLCT-inferred velocities obtained from $B_z$ serve as an excellent reference for expected results when applying FLCT to Stokes parameters in a spectral line. The following section examines tracking on $B_z$ inferred from synthetic TuMag observations.

\section{Application to synthetic polarized spectra}
\label{sec:tumag}

The previous section demonstrated that applying FLCT to $B_z$ magnetograms \text{extracted from simulation data} yields a reliable inference of horizontal velocities for atmospheric layers from $\log\tau_{500}=0$ to $\log\tau_{500}=-4$ (see Figs.\,\ref{fig:track_BzTemp} and \ref{fig:track_BzTemp300}, and Table \ref{table:2}). Next, we applied FLCT to the magnetograms obtained from the synthetic TuMag observations described in Sect. \ref{ssec:synt}. By tracking the magnetograms obtained from the Fe\,I\,525.0\,nm and Mg\,I\,b2 lines, we aimed to track the horizontal velocity at the atmospheric heights probed by these two lines.

\subsection{Inference of $B_z$}
\label{ssec:inversions}
To infer $B_z$ from the Stokes spectra of the Fe\,I\,525.0 line, we used pyMilne, an efficient, parallel, spatially coupled Milne-Eddington inversion code \citep{delaCruz2019pymilne}. The Milne-Eddington approximation assumes a linear variation of the source function with depth, while the magnetic field and LOS velocity remain constant with depth. We applied pyMilne to each individual synthetic Stokes cube $\vec{I}(x,y, \lambda)$, where \vec{I} denotes the Stokes vector,  $\vec{I} = (I,Q,U,V)^T$. Each inversion yields a vector magnetogram $\vec{B}(x,y)$ for a given time step. It is important to adopt a so-called ``formation height'' for the line, to ascribe this inferred magnetic field to a specific atmospheric layer. To this end, we compared the inferred $B_z$ to the original $B_z$ at different iso-$\tau_{500}$ surfaces \citep[by calculating the cross-correlation between the two, following, e.g.][]{Vukadinovic2022mg}. We find the best agreement for $\log \tau_{500} = -1$, which is consistent with the typical formation heights of moderately strong photospheric lines \citep[see, e.g.,][]{borrero2014me}. 

The Mg\,I\,b2 line is significantly stronger than Fe\,I\,525.0\,nm, which makes the Milne-Eddington approximation invalid \citep[but see ][]{Dorantes2022modified}. We therefore used the weak-field approximation (WFA) to infer the longitudinal field, which coincides with $B_z$, under the disk-center assumption and in the absence of magnetic substructures below the pixel scale. The weak-field approximation assumes that the magnetic field is constant with height and that the Zeeman splitting it produces is much smaller than other sources of line broadening. This leads to the following relationship \citep[e.g.][]{LL2004bible}:
\begin{equation}
    V_\lambda = -4.67\times 10^{-13} \frac{dI_\lambda}{d\lambda}\lambda_0^2 g_L B_z,
\end{equation}
where the wavelengths are in $\AA$, magnetic field is in G, and $g_L$ is the effective Land\'{e} factor of the line, in this case $g_L = 1.75$. Given the large width of the Mg\,I line, the weak-field approximation is valid and probes depths around $\log\tau_{500}=-3.3$  \citep{Vukadinovic2022mg}. We confirmed this by applying the weak-field approximation to the core of the Mg\,I\ line (0.2\,$\AA$ around the intensity minimum) and comparing the inferred magnetic field with the original one in the simulation, again using cross-correlation. The best agreement occurs between $\log\tau_{500}=-3$ and $\log\tau_{500}=-4$. 

Therefore, the two spectral lines probe different atmospheric layers: the mid-photosphere for Fe\,I\,525.0\,nm and the upper-photosphere or temperature-minimum region for Mg\,I\,b2. This is also evident in Fig.\,\ref{fig:multi_lambda_line}, where the Stokes $V$ map in the Mg\,I line shows significantly less structure and more diffuse circular-polarization signals. 

\subsection{Tracking synthetic magnetograms}

\begin{figure}
    \centering
    \includegraphics[width=0.495\textwidth]{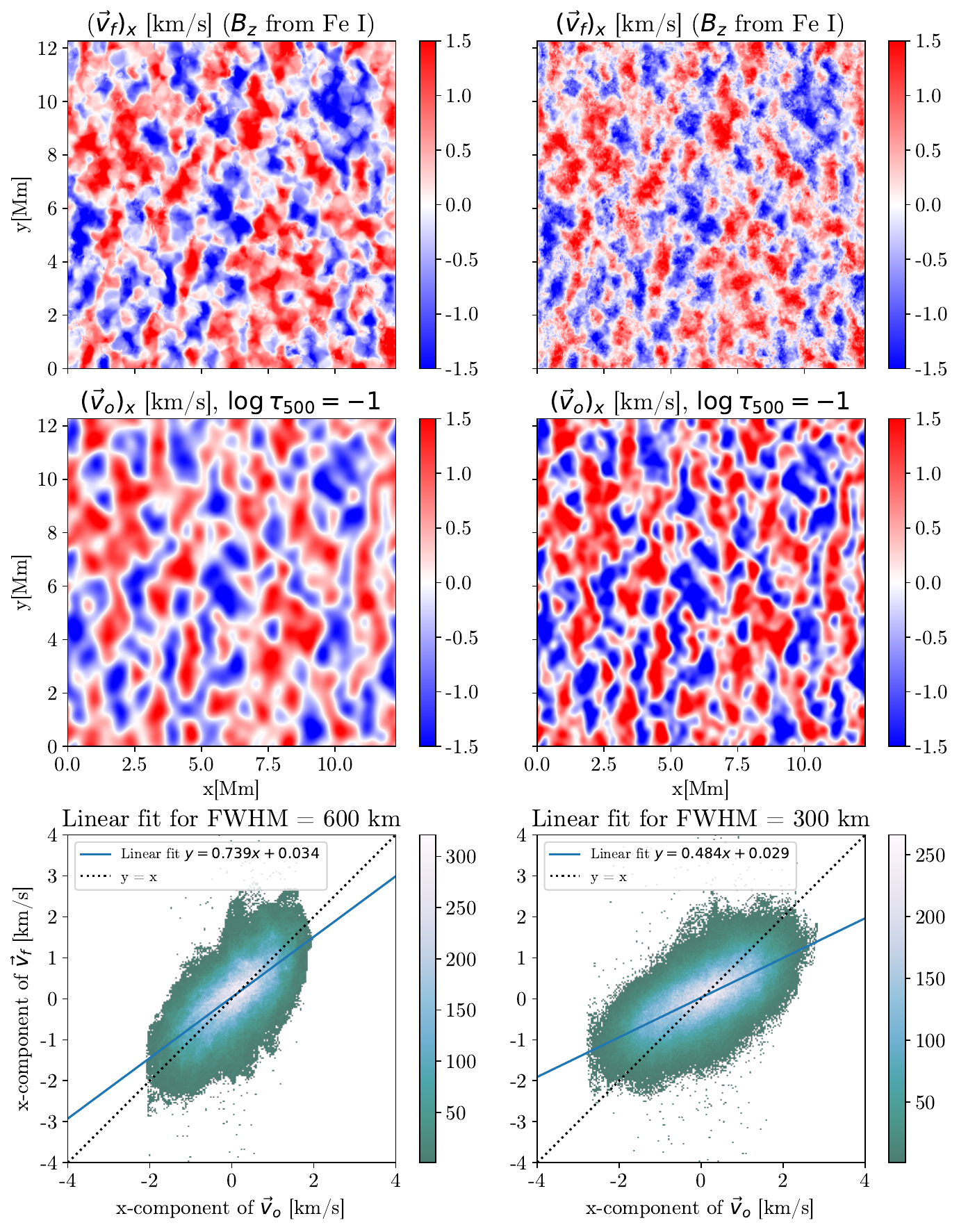}
    \caption{Comparison between the $x$ component of the horizontal velocity retrieved by FLCT applied to the synthetic magnetogram inferred from the Fe\,I line and the simulated velocity at $\log\tau_{500}=-1$.} 
    \label{fig:track_Bz_FeI}
\end{figure}

\begin{figure}
    \centering
    \includegraphics[width=0.495\textwidth]{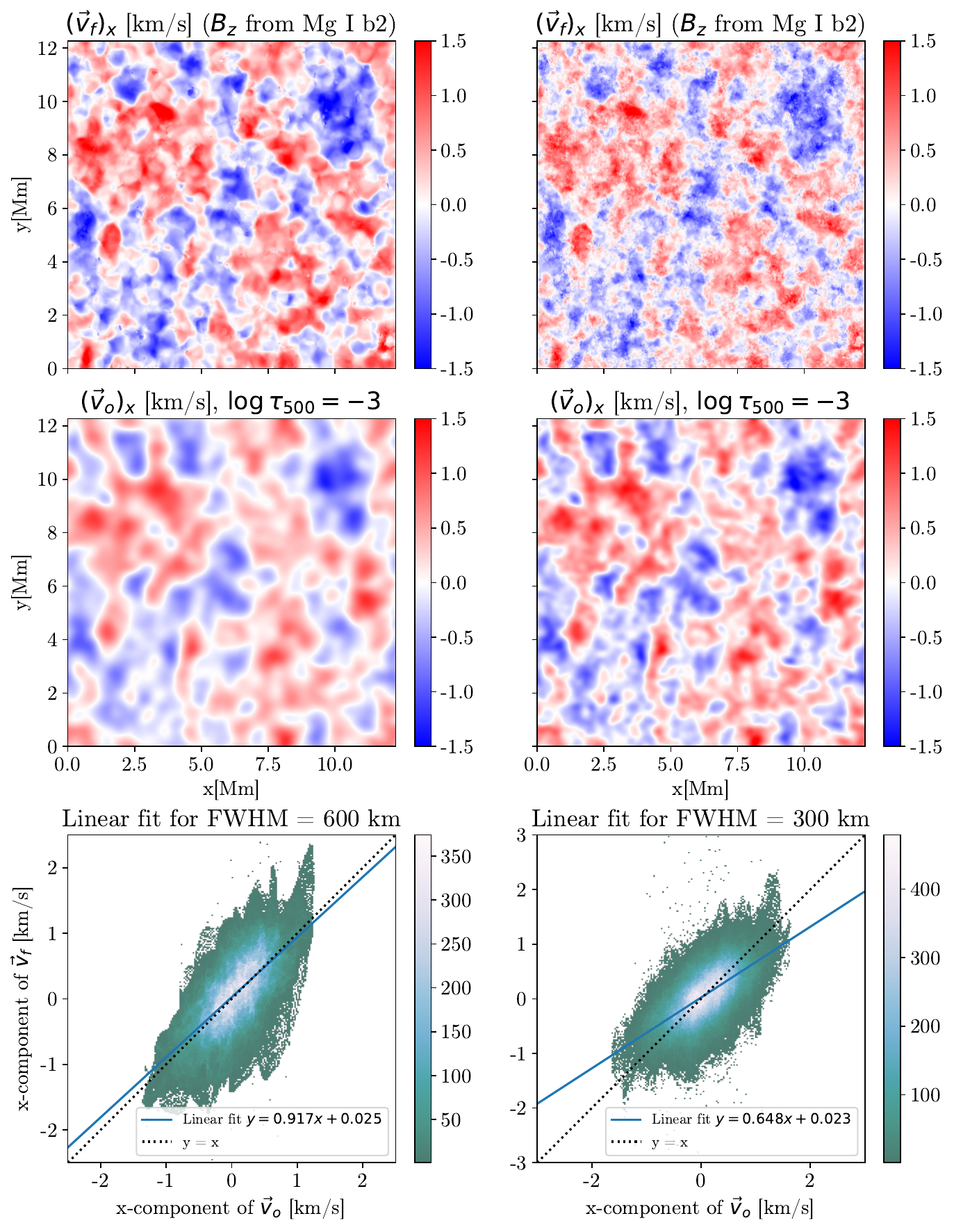}
    \caption{Same as Fig.\,\ref{fig:track_Bz_FeI}, but showing the synthetic magnetograms inferred from the Mg\,I line compared with the simulated velocity at $\log\tau_{500}=-3$.} 
    \label{fig:track_Bz_MgI}
\end{figure}

\begin{figure}
    \centering
    \includegraphics[width=0.495\textwidth]{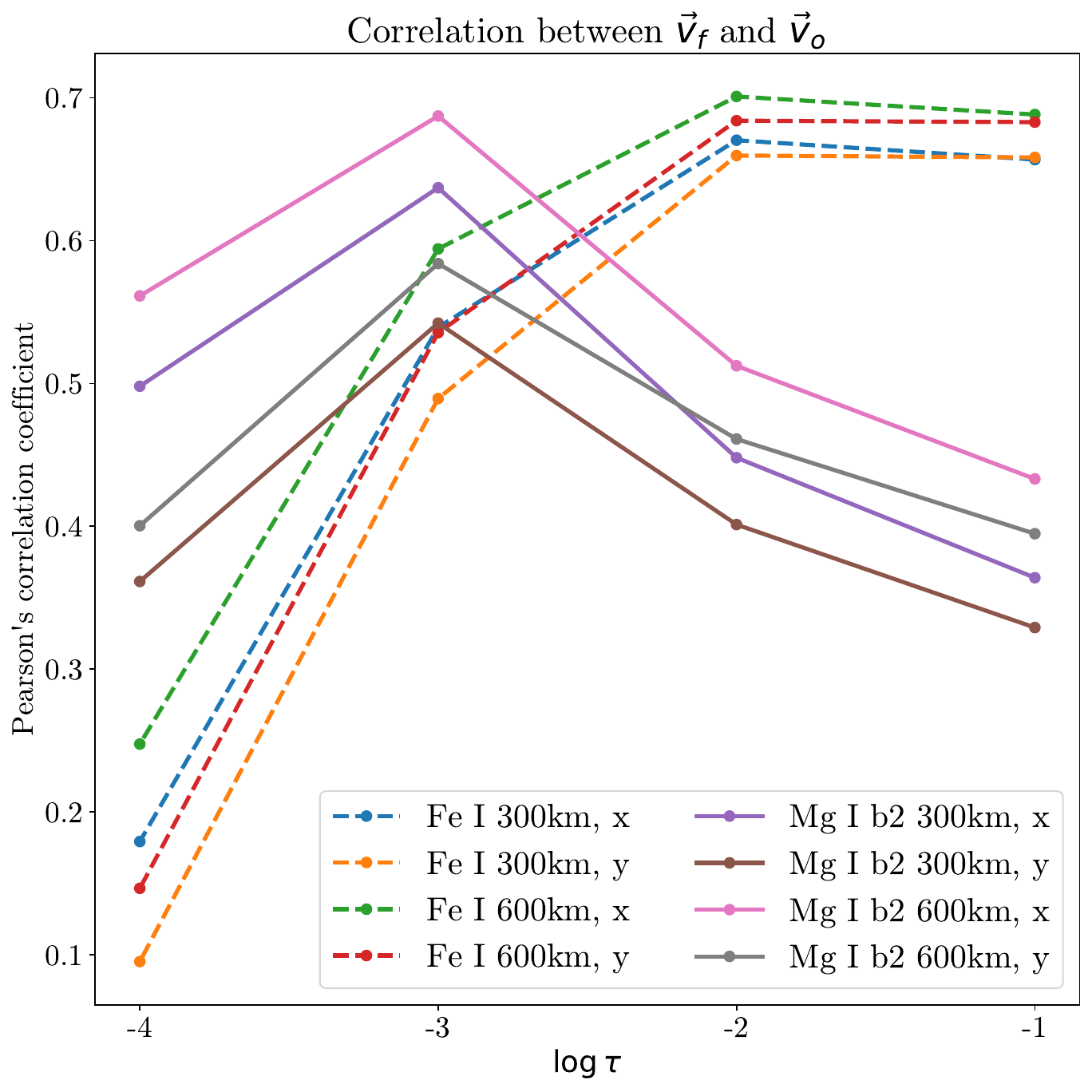}
    \caption{Pearson correlation coefficients between simulation velocities and velocities recovered by FLCT from $B_z$ inferred either from Milne-Eddington inversions of the Fe\,I\,$\SI{525.0}{nm}$ line or from $B_z$ inferred using the WFA approximation with the Mg I b2 line at $\log\tau_{500}=\{-1, -2, -3, -4\}$, respectively. Correlations are shown only for filtered simulation velocities.}
    \label{fig:track_tables}
\end{figure}

We applied FLCT to the synthetic magnetograms obtained from the two spectral lines to infer $\vec{v}_f$ and compared it with $\vec{v}_o$ at various optical depths. The synthetic spectra and magnetograms have different spatial sampling compared to the original simulation (32\,km compared to 16\,km). Figures \ref{fig:track_Bz_FeI} and \ref{fig:track_Bz_MgI} show the agreement between the inferred and the original velocities for two specific heights: $\log\tau_{500}=-1$ for the Fe line and $\log\tau_{500}=-3$ for the Mg line. Fig.\,\ref{fig:track_tables} summarizes the comparison between the inferred and the original velocities. The $\vec{v}_f$ inferred from the Fe\,I\,525.0\,nm line shows good agreement with $\vec{v}_o(\log\tau_{500}=-1)$ and $\vec{v}_o(\log\tau_{500}=-2)$, but the correlation decreases rapidly toward the higher layers. This behavior is consistent with our estimate of the depth where the Fe\,I line probes the magnetic field. The velocities inferred from the Mg\,I\,b2 line magnetograms agree best with $\vec{v}_o(\log\tau_{500}=-3)$ and $\vec{v}_o(\log\tau_{500}=-4)$, thereby probing regions near the temperature minimum and agreeing closely with the estimates of \citet{Vukadinovic2022mg}. Our results suggest that these two lines complement each other effectively in covering the whole vertical extent from the base of the photosphere (continuum) through the mid-photosphere (Fe\,I line) up to the temperature-minimum region (Mg\,I line).

The lower panels of Figs.\,\ref{fig:track_Bz_FeI} and \ref{fig:track_Bz_MgI} show that velocities inferred by tracking the synthetic magnetograms are also underestimated. This underestimation is larger for a smaller apodizing window, as already shown in Figs.\,\ref{fig:track_I} and \ref{fig:track_B}. Furthermore, tracking synthetic magnetograms in the Mg\,I\,b2 line yields velocity amplitudes closer to the simulated values than those obtained from the Fe\,I line. This likely occurs because this line traces velocities higher in the atmosphere, where the flows are more diffuse and less spatially structured. Consequently, the FLCT algorithm can capture the properties of the flow more accurately.

For the spectral lines considered, applying tracking to the synthetic magnetograms yields somewhat poorer agreement than direct tracking of the magnetic fields at fixed optical depths (see Sect. \ref{ssec::track_multi}). This outcome is not unexpected. The magnetic field in the solar atmosphere is height-dependent, and thus both the Milne-Eddington and WFA inversions introduce noticeable discrepancies and systematic errors. Depth-dependent magnetic field diagnostics using a more sophisticated inversion code could yield improvement, but at a significant computational cost, particularly due to the NLTE treatment of the Mg\,I\,b2 line \citep[see the recent results of][]{SiuTapia2025MgIb1, SiuTapia2025MgIb2}.

\subsection{Velocity divergence inferred from synthetic magnetograms}

To further assess the reliability of horizontal velocity diagnostics using these spectral lines, we compared the divergence calculated from 
$\vec{v}_f$, inferred from synthetic magnetograms of the Fe\,I and Mg\,I spectral lines, with the value calculated from $\vec{v}_o$. In this context, the horizontal divergence is defined as 
\begin{equation}
    \nabla_h \cdot \vec{v} = \frac{\partial v_x}{\partial x} + \frac{\partial v_y}{\partial y}.
\end{equation}
For instance, granules are expected to exhibit positive divergence, whereas intergranular lanes are dominated by negative divergence, corresponding to regions where flows converge.

The agreement is significantly poorer than for the individual comparison of the $v_x$ and $v_y$ components. For the magnetograms inferred from the Fe\,I line, the correlation coefficient is $0.29$ for a 600 km FWHM and $0.20$ for a 300 km FWHM. The agreement is even poorer for the Mg\,I\,b2 line, with  correlation coefficients of $0.23$ and $0.11$ for the two apodizing windows, respectively. 
The poor estimation of divergences may stem from the fact that FLCT estimates velocities in neighboring pixels independently and does not incorporate spatial variations in the flow structure when estimating each pixel's flow \citep[cf. the DAVE, which is short for differential affine velocity estimator, method described by][]{Schuck_2006_DAVE}. Consequently, no correlation is enforced between the neighboring pixel velocities. However, based on the nature of measurement uncertainties, the numerical differentiation used to compute the flow divergence also contributes to the discrepancy, as the uncertainty in the estimated velocity differences exceeds that of the individual velocity estimates. Consequently, we expect the derivatives of the estimated flows to be noisier than the flows themselves. Another source of disagreement arises because the simplified inversion methods effectively probe different depths at different pixels, which results in a different corrugation of inferred magnetic field maps between the original iso-$\tau_{500}$ slices and the obtained synthetic magnetograms.

\begin{figure}
    \centering
    \includegraphics[width=0.495\textwidth]{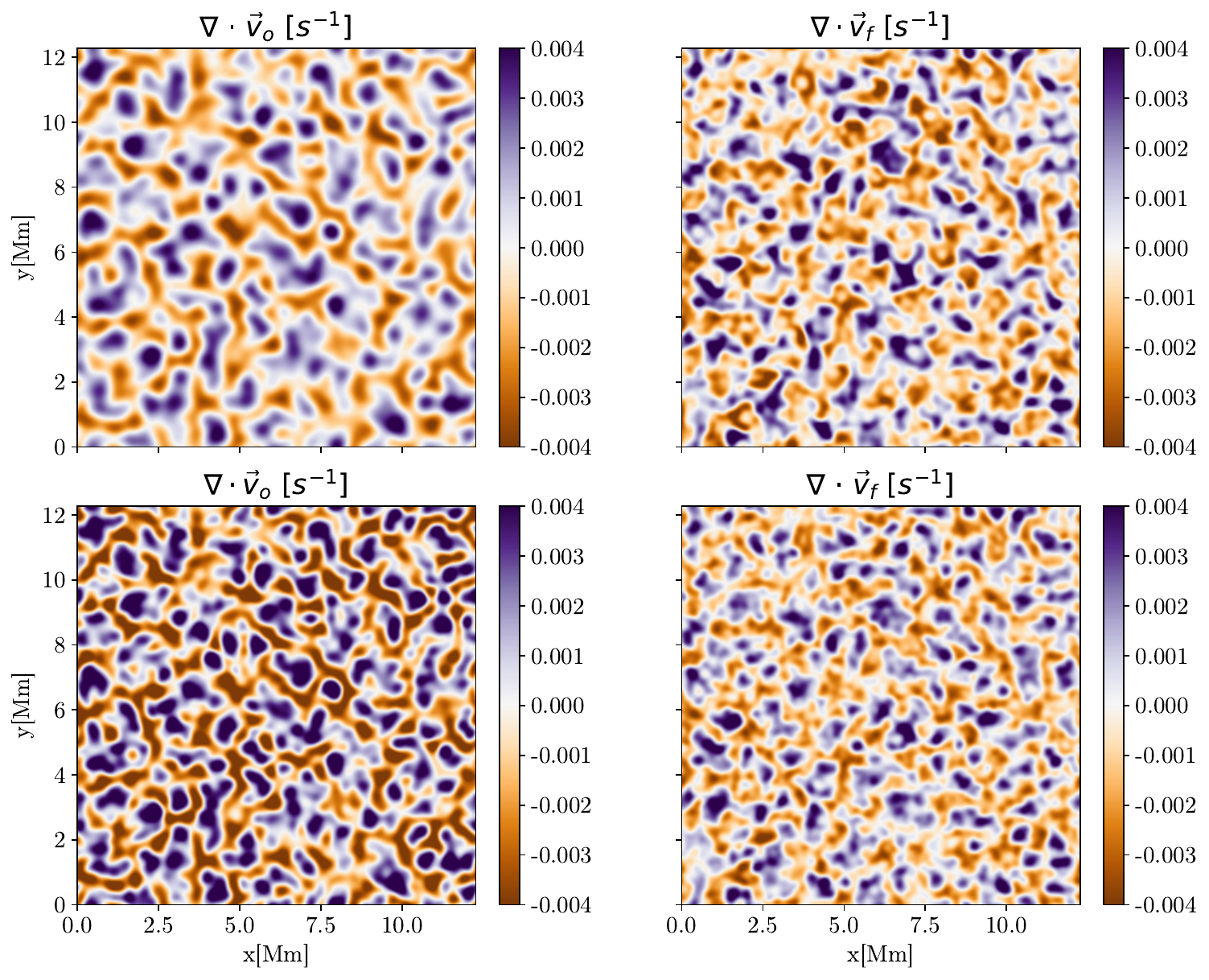}
    \caption{Comparison between velocity divergence calculated using $\vec{v}_o$ (left panels) at $\log \tau_{500} = -1$ and $\vec{v}_f$ inferred from the Fe\,I line (right panels). Top: Results with an apodizing window of 600\,km. Bottom: Results with an apodizing window of 300\,km.}
    \label{fig:Fe_I_div}
\end{figure}

\begin{figure}
    \centering
    \includegraphics[width=0.495\textwidth]{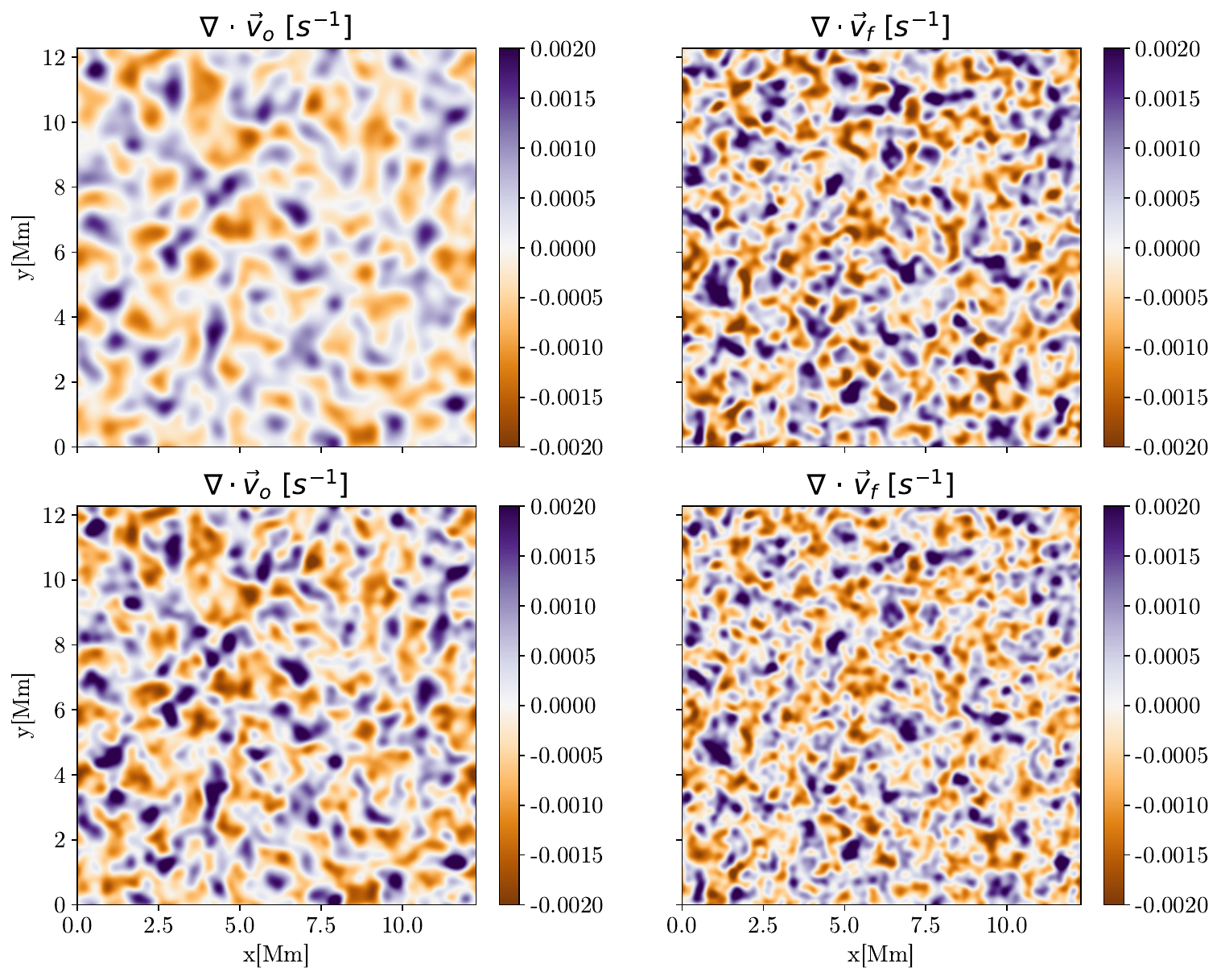}
    \caption{Same as Fig.\,\ref{fig:Fe_I_div} but for the Mg\,I line and $\log\tau_{500} = -3$.}
    \label{fig:Mg_I_div}
\end{figure}

To improve the agreement, we smoothed both the original and inferred divergence maps with a Gaussian filter of $\sigma = 200\,$km, thereby further reducing the effective spatial resolution. This smoothing significantly improved the correlation coefficients to $\approx$ 0.6 for the Fe\,I magnetograms and $\approx$ 0.45 for the Mg\,I magnetograms (see Table\,\ref{table:6}). The comparison between the divergence inferred from the Fe\,I line and that derived from $\vec{v}_o$ at $\log\tau_{500}=-1$ is presented in Fig.\,\ref{fig:Fe_I_div}, while Fig.\,\ref{fig:Mg_I_div} shows the same comparison, but for the Mg\,I line and $\vec{v}_o$ at $\log\tau_{500}=-3$. The pronounced difference between the ability of FLCT to reproduce horizontal velocities and their divergence indicates the presence of systematic discrepancies in the FLCT inference of horizontal velocities, as previously reported by \citet{Gibson2004_rope_emerging} and \citet{ERempel2022tracking}. This limitation of the FLCT technique is expected to be improved by recently developed machine learning techniques \citep[e.g.][]{aar_deepvel, Lennard_2025_DeepVel}.

\begin{table}[h!]
\caption{Pearson correlation coefficient for divergence of FLCT and simulated velocities.}
\begin{center}
\begin{adjustbox}{width=0.48\textwidth}
  \begin{tabular}{ @{} lcSSSSS @{} }
    \toprule
    \multirow{2}{*}{Div.} &
    \multirow{2}{*}{FWHM[km]} &
      \multicolumn{1}{c}{$\log\tau_{500} = -1$} &
      \multicolumn{1}{c}{$\log\tau_{500} = -2$} &
      \multicolumn{1}{c}{$\log\tau_{500} = -3$} &
      \multicolumn{1}{c}{$\log\tau_{500} = -4$} \\
      && {$r$} & {$r$} & {$r$} & {$r$} \\
      \midrule
     \multirow{2}{*}{Fe} & 600 & 0.62 & 0.62 & 0.51 & -0.05 \\
     & 300 & 0.64 & 0.65 & 0.48 & -0.08 \\
      \cmidrule(l){2-6} 
    \multirow{2}{*}{Mg} & 600 & 0.29 & 0.34 & 0.46 & 0.21 \\
    & 300 & 0.27 & 0.33 & 0.47 & 0.20 \\
    \bottomrule
  \end{tabular}
  
    \label{table:6}
\end{adjustbox}
  \end{center}
  \tablefoot{The FLCT velocities are recovered from $B_z$ inferred either from Milne-Eddington inversion of Fe\,I\,$\SI{525.0}{nm}$ or from the WFA approximation using the Mg I b2 line and simulation velocities at $\log\tau_{500}=\{-1, -2, -3, -4\}$, after applying an additional Gaussian filter. Correlation is shown only for filtered simulation velocities.}
\end{table}

\section{Conclusions}
\label{sec:conclusions}

We have studied the diagnostic potential of the FLCT flow-tracking technique to probe horizontal plasma flows at multiple heights in the lower solar atmosphere. First, we applied FCLT to a time series of intensitygrams, magnetograms, and temperature maps at constant optical depth surfaces ranging from $\log\tau_{500}=0$ to $\log\tau_{500}=-4$, directly obtained from the simulation. After spatial filtering and temporal averaging (which is a standard procedure for velocities obtained with optical flow methods), we conclude that tracking LOS magnetograms yields satisfactory agreement with the original velocities across all considered optical depths. In contrast, tracking the temperature is unreliable for atmospheric layers above $\log\tau_{500}=-1$.

Next, we tracked synthetic magnetograms obtained through a simple but robust inversion of synthetic spectropolarimetric observables in two spectral lines: Fe\,I\,525.0\,nm and Mg\,I\,b2. These two lines probe the low-to-mid photosphere and the upper-photosphere or temperature-minimum regions, respectively. We synthesized the polarized spectra of the two lines using the SNAPI code \citep{Milic2018snapi} and then obtained synthetic magnetograms using the Milne-Eddington inversion code pyMilne \citep{delaCruz2019pymilne} for the Fe\,I line and the weak-field approximation for the Mg\,I line. The inferred velocities agree well with the suitably filtered original ones, achieving correlation coefficients of $r=0.65-0.7$. This result holds for both apodizing windows considered: 300\,km and 600\,km. The vertical magnetic field inferred from the Fe\,I\,525.0\,nm line accurately tracks horizontal velocities between $\log\tau_{500}=-1$ and $\log\tau_{500}=-2$, whereas the field inferred from the Mg\,I\,b2 line tracks the velocities in the higher regions, from $\log\tau_{500}=-3$ to $\log\tau_{500}=-4$. Therefore, fast-cadence imaging spectropolarimetry in these two spectral lines enables a robust inference of the horizontal velocities from the base of the photosphere to the temperature minimum. The important advantage of using spectral lines to track plasma flows is the potential to infer the vertical velocity or its vertical gradient \citep[e.g.,][]{Michiel_Jaime_2017}. However, the FLCT technique performs significantly worse when calculating the horizontal divergence of the velocity, yielding correlation coefficients of 0.6 or less. 

These findings are especially relevant for the observations collected by the SUNRISE III/TuMag \citep{TuMag}, the newly upgraded CRISP \citep{CRISP2008}, and the upcoming VTF \citep{vtf} instruments. The robust and simple magnetic field inference proposed here should be suitable for the relatively coarse wavelength sampling offered by these Fabry-Perot-based spectropolarimeters. The application to data from integral field units (IFUs) such as MiHI \citep{vanNoort2022MiHI} is more complex due to their spectral resolution, which may require more comprehensive inversion approaches. 
The limited range of spatial scales probed by FLCT represents a significant limitation of this method.
Our results exhibit good agreement for apodizing windows of 600\,km and 300\,km, which are well above the diffraction limit of the instruments considered ($\approx80$\,km for SST and SUNRISE, and $\approx20\,$km for DKIST, at visible wavelengths). 

Inference of horizontal velocities from high-resolution observations requires further development and extensions of FLCT techniques or the application of recently developed AI-based methods, such as DeepVel \citep{aar_deepvel}. Given that DeepVel can, in principle, infer horizontal velocities at multiple heights, training and applying it to spectral line data is expected to significantly improve its performance. Finally, the development of techniques that couple spectropolarimetric diagnostics with MHD equations \citep{borrero_mhs_I, borrero_mhs_IV}, possibly combined with physically informed neural networks \citep{Jarolim2023pinn, Tremblay2023}, is expected to provide the most robust and physically meaningful information on plasma conditions, including horizontal velocities.

\begin{acknowledgement}
We gratefully acknowledge the ISSI team devoted to tracking flows in the solar atmosphere. We are indebted to an anonymous referee for their careful reading of the paper and insightful comments, which significantly improved the clarity and presentation of this work.

TK and IM acknowledge the financial support from the Serbian Ministry of Science and Technology through the grants 451-03-136/2025-03/200104 and 451-03-136/2025-03/200002.
M.D.K.\ acknowledges support by NASA ECIP award 80NSSC19K0910, NASA HSR-80NSSC23K0092 and NSF CAREER award SPVKK1RC2MZ3.
AAR acknowledges support from the Agencia Estatal de Investigaci\'on del 
Ministerio de Ciencia, Innovaci\'on y Universidades (MCIU/AEI) 
and the European Regional Development Fund (ERDF) through project PID2022-136563NB-I0.
BTW gratefully acknowledges support from NASA HSR-80NSSC23K0092. 
This material is based upon work supported by the NSF National Center for Atmospheric Research, which is a major facility sponsored by the U.S. National Science Foundation under Cooperative Agreement No. 1852977. We would like to acknowledge high-performance computing support from the Derecho system (doi:10.5065/qx9a-pg09) provided by the NSF National Center for Atmospheric Research (NCAR), sponsored by the National Science Foundation.
This research has made use of NASA's Astrophysics Data System.  
\end{acknowledgement}
\bibliography{tracking_lines}

@ARTICLE{aar_deepvel,
       author = {{Asensio Ramos}, A. and {Requerey}, I.~S. and {Vitas}, N.},
        title = "{DeepVel: Deep learning for the estimation of horizontal velocities at the solar surface}",
      journal = {\aap},
         year = 2017,
        month = jul,
       volume = {604},
          eid = {A11},
        pages = {A11},
          doi = {10.1051/0004-6361/201730783},
archivePrefix = {arXiv},
       eprint = {1703.05128},
 primaryClass = {astro-ph.SR},
       adsurl = {https://ui.adsabs.harvard.edu/abs/2017A&A...604A..11A}
}

@ARTICLE{Afanasyev2021,
       author = {{Afanasyev}, Andrey N. and {Kazachenko}, Maria D. and {Fan}, Yuhong and {Fisher}, George H. and {Tremblay}, Benoit},
        title = "{Validation of the PDFI\_SS Method for Electric Field Inversions Using a Magnetic Flux Emergence Simulation}",
      journal = {\apj},
         year = 2021,
        month = sep,
       volume = {919},
       number = {1},
          eid = {7},
        pages = {7},
          doi = {10.3847/1538-4357/ac0d01},
archivePrefix = {arXiv},
       eprint = {2106.10579},
 primaryClass = {astro-ph.SR},
       adsurl = {https://ui.adsabs.harvard.edu/abs/2021ApJ...919....7A}
}

@ARTICLE{arregui2015,
       author = {{Arregui}, I{\~n}igo},
        title = "{Wave heating of the solar atmosphere}",
      journal = {Philos. Trans. R. Soc. A},
         year = 2015,
        month = apr,
       volume = {373},
       number = {2042},
        pages = {20140261-20140261},
          doi = {10.1098/rsta.2014.0261},
archivePrefix = {arXiv},
       eprint = {1501.06708},
 primaryClass = {astro-ph.SR},
       adsurl = {https://ui.adsabs.harvard.edu/abs/2015RSPTA.37340261A}
}

@ARTICLE{bendza2018,
       author = {{Tremblay}, Benoit and {Roudier}, Thierry and {Rieutord}, Michel and {Vincent}, Alain},
        title = "{Reconstruction of Horizontal Plasma Motions at the Photosphere from Intensitygrams: A Comparison Between DeepVel, LCT, FLCT, and CST}",
      journal = {\solphys},
         year = 2018,
        month = apr,
       volume = {293},
       number = {4},
          eid = {57},
        pages = {57},
          doi = {10.1007/s11207-018-1276-7},
       adsurl = {https://ui.adsabs.harvard.edu/abs/2018SoPh..293...57T}
}

@ARTICLE{borrero_mhs_I,
       author = {{Borrero}, J.~M. and {Pastor Yabar}, A. and {Rempel}, M. and {Ruiz Cobo}, B.},
        title = "{Combining magnetohydrostatic constraints with Stokes profiles inversions. I. Role of boundary conditions}",
      journal = {\aap},
         year = 2019,
        month = dec,
       volume = {632},
          eid = {A111},
        pages = {A111},
          doi = {10.1051/0004-6361/201936367},
archivePrefix = {arXiv},
       eprint = {1910.14131},
 primaryClass = {astro-ph.SR},
       adsurl = {https://ui.adsabs.harvard.edu/abs/2019A&A...632A.111B}
}

@ARTICLE{borrero_mhs_IV,
       author = {{Borrero}, J.~M. and {Pastor Yabar}, A. and {Ruiz Cobo}, B.},
        title = "{Combining magneto-hydrostatic constraints with Stokes profile inversions. IV. Imposing {\ensuremath{\nabla}}{\ensuremath{\cdot}}B = 0 condition}",
      journal = {\aap},
         year = 2024,
        month = jul,
       volume = {687},
          eid = {A155},
        pages = {A155},
          doi = {10.1051/0004-6361/202449572},
archivePrefix = {arXiv},
       eprint = {2404.02045},
 primaryClass = {astro-ph.SR},
       adsurl = {https://ui.adsabs.harvard.edu/abs/2024A&A...687A.155B}
}

@ARTICLE{borrero2014me,
       author = {{Borrero}, J.~M. and {Lites}, B.~W. and {Lagg}, A. and {Rezaei}, R. and {Rempel}, M.},
        title = "{Comparison of inversion codes for polarized line formation in MHD simulations. I. Milne-Eddington codes}",
      journal = {\aap},
         year = 2014,
        month = dec,
       volume = {572},
          eid = {A54},
        pages = {A54},
          doi = {10.1051/0004-6361/201424584},
archivePrefix = {arXiv},
       eprint = {1409.3376},
 primaryClass = {astro-ph.SR},
       adsurl = {https://ui.adsabs.harvard.edu/abs/2014A&A...572A..54B}
}

@ARTICLE{Chae2008,
       author = {{Chae}, Jongchul and {Sakurai}, Takashi},
        title = "{A Test of Three Optical Flow Techniques{\textemdash}LCT, DAVE, and NAVE}",
      journal = {\apj},
         year = 2008,
        month = dec,
       volume = {689},
       number = {1},
        pages = {593-612},
          doi = {10.1086/592761},
       adsurl = {https://ui.adsabs.harvard.edu/abs/2008ApJ...689..593C}
}

@ARTICLE{CRISP2008,
       author = {{Scharmer}, G.~B. and {Narayan}, G. and {Hillberg}, T. and {de la Cruz Rodriguez}, J. and {L{\"o}fdahl}, M.~G. and {Kiselman}, D. and {S{\"u}tterlin}, P. and {van Noort}, M. and {Lagg}, A.},
        title = "{CRISP Spectropolarimetric Imaging of Penumbral Fine Structure}",
      journal = {\apjl},
         year = 2008,
        month = dec,
       volume = {689},
       number = {1},
        pages = {L69},
          doi = {10.1086/595744},
archivePrefix = {arXiv},
       eprint = {0806.1638},
 primaryClass = {astro-ph},
       adsurl = {https://ui.adsabs.harvard.edu/abs/2008ApJ...689L..69S}
}

@ARTICLE{delaCruz2019pymilne,
       author = {{de la Cruz Rodr{\'\i}guez}, J.},
        title = "{A method for global inversion of multi-resolution solar data}",
      journal = {\aap},
         year = 2019,
        month = nov,
       volume = {631},
          eid = {A153},
        pages = {A153},
          doi = {10.1051/0004-6361/201936635},
archivePrefix = {arXiv},
       eprint = {1909.02604},
 primaryClass = {astro-ph.SR},
       adsurl = {https://ui.adsabs.harvard.edu/abs/2019A&A...631A.153D}
}

@article{Demoulin_Berger_2003,
author = {Demoulin, Pascal and Berger, Mitchell},
year = {2003},
month = {01},
pages = {203-215},
title = {Magnetic Energy and Helicity Fluxes at the Photospheric Level},
volume = {215},
journal = {\solphys},
doi = {10.1023/A:1025679813955}
}

@ARTICLE{Dorantes2022modified,
       author = {{Dorantes-Monteagudo}, A.~J. and {Siu-Tapia}, A.~L. and {Quintero-Noda}, C. and {Orozco Su{\'a}rez}, D.},
        title = "{A modified Milne-Eddington approximation for a qualitative interpretation of chromospheric spectral lines}",
      journal = {\aap},
         year = 2022,
        month = mar,
       volume = {659},
          eid = {A156},
        pages = {A156},
          doi = {10.1051/0004-6361/202142810},
archivePrefix = {arXiv},
       eprint = {2112.14536},
 primaryClass = {astro-ph.SR},
       adsurl = {https://ui.adsabs.harvard.edu/abs/2022A&A...659A.156D}
}

@ARTICLE{Pauldynamo2010,
       author = {{Charbonneau}, Paul},
        title = "{Dynamo Models of the Solar Cycle}",
      journal = {Living Rev. Sol. Phys.},
         year = 2010,
        month = dec,
       volume = {7},
       number = {1},
          eid = {3},
        pages = {3},
          doi = {10.12942/lrsp-2010-3},
       adsurl = {https://ui.adsabs.harvard.edu/abs/2010LRSP....7....3C}
}

@article{Fisher_2020,
doi = {10.3847/1538-4365/ab8303},
url = {https://dx.doi.org/10.3847/1538-4365/ab8303},
year = {2020},
month = {apr},
publisher = {The American Astronomical Society},
volume = {248},
number = {1},
pages = {2},
author = {George H. Fisher and Maria D. Kazachenko and Brian T. Welsch and Xudong Sun and Erkka Lumme and David J. Bercik and Marc L. DeRosa and Mark C. M. Cheung},
title = {The PDFI_SS Electric Field Inversion Software},
journal = {\apjs}
}

@ARTICLE{Iniesta2016rev,
       author = {{del Toro Iniesta}, Jose Carlos and {Ruiz Cobo}, Basilio},
        title = "{Inversion of the radiative transfer equation for polarized light}",
      journal = {Living Rev. Sol. Phys.},
         year = 2016,
        month = dec,
       volume = {13},
       number = {1},
          eid = {4},
        pages = {4},
          doi = {10.1007/s41116-016-0005-2},
archivePrefix = {arXiv},
       eprint = {1610.10039},
 primaryClass = {astro-ph.SR},
       adsurl = {https://ui.adsabs.harvard.edu/abs/2016LRSP...13....4D}
}

@ARTICLE{Gibson2004_rope_emerging,
       author = {{Gibson}, S.~E. and {Fan}, Y. and {Mandrini}, C. and {Fisher}, G. and {Demoulin}, P.},
        title = "{Observational Consequences of a Magnetic Flux Rope Emerging into the Corona}",
      journal = {\apj},
         year = 2004,
        month = dec,
       volume = {617},
       number = {1},
        pages = {600-613},
          doi = {10.1086/425294},
       adsurl = {https://ui.adsabs.harvard.edu/abs/2004ApJ...617..600G}
}

@ARTICLE{Kazachenko2014,
   author = {{Kazachenko}, M.~D. and {Fisher}, G.~H. and {Welsch}, B.~T.},
    title = "{A Comprehensive Method of Estimating Electric Fields from Vector Magnetic Field and Doppler Measurements}",
  journal = {\apj},
archivePrefix = "arXiv",
   eprint = {1404.4027},
 primaryClass = "astro-ph.SR",
     year = 2014,
    month = nov,
   volume = 795,
      eid = {17},
    pages = {17},
      doi = {10.1088/0004-637X/795/1/17},
   adsurl = {http://adsabs.harvard.edu/abs/2014ApJ...795...17K}
}

@ARTICLE{Lennard_2025_DeepVel,
       author = {{Lennard}, Matthew G. and {Silva}, Suzana S.~A. and {Tremblay}, Benoit and {Ramos}, Andr{\'e}s Asensio and {Verth}, Gary and {Ballai}, Istvan and {Iijima}, Haruhisa and {Hotta}, Hideyuki and {Rempel}, Matthias and {Park}, Sung-Hong and {Fedun}, Viktor},
        title = "{Recovering coherent flow structures in active regions using machine learning}",
      journal = {\mnras},
         year = 2025,
        month = may,
       volume = {539},
       number = {3},
        pages = {2498-2512},
          doi = {10.1093/mnras/staf576},
       adsurl = {https://ui.adsabs.harvard.edu/abs/2025MNRAS.539.2498L}
}

@article{Li_2021,
doi = {10.3847/1538-4357/ac043e},
url = {https://dx.doi.org/10.3847/1538-4357/ac043e},
year = {2021},
month = {jul},
publisher = {The American Astronomical Society},
volume = {915},
number = {2},
pages = {124},
author = {Z. F. Li and X. Cheng and M. D. Ding and Katharine K. Reeves and DeOndre Kittrell and Mark Weber and David E. McKenzie},
title = {Thermodynamic Evolution of Solar Flare Supra-arcade Downflows},
journal = {\apj}
}

@BOOK{LL2004bible,
       author = {{Landi Degl'Innocenti}, E. and {Landolfi}, M.},
        title = "{Polarization in Spectral Lines}",
         year = 2004,
       volume = {307},
          doi = {10.1007/978-1-4020-2415-3},
       adsurl = {https://ui.adsabs.harvard.edu/abs/2004ASSL..307.....L}
}

@ARTICLE{Loptien2016,
       author = {{L{\"o}ptien}, B. and {Birch}, A.~C. and {Duvall}, T.~L. and {Gizon}, L. and {Schou}, J.},
        title = "{The shrinking Sun: A systematic error in local correlation tracking of solar granulation}",
      journal = {\aap},
         year = 2016,
        month = may,
       volume = {590},
          eid = {A130},
        pages = {A130},
          doi = {10.1051/0004-6361/201628112},
archivePrefix = {arXiv},
       eprint = {1604.04469},
 primaryClass = {astro-ph.SR},
       adsurl = {https://ui.adsabs.harvard.edu/abs/2016A&A...590A.130L}
}

@Article{Lumme2019,
author="Lumme, E.
and Kazachenko, M. D.
and Fisher, G. H.
and Welsch, B. T.
and Pomoell, J.
and Kilpua, E. K. J.",
title="Probing the Effect of Cadence on the Estimates of Photospheric Energy and Helicity Injections in Eruptive Active Region NOAA AR 11158",
journal="\solphys",
year="2019",
month="Jun",
day="27",
volume="294",
number="6",
pages="84",
issn="1573-093X",
doi="10.1007/s11207-019-1475-x",
url="https://doi.org/10.1007/s11207-019-1475-x"
}

@ARTICLE{Milic2018snapi,
       author = {{Mili{\'c}}, I. and {van Noort}, M.},
        title = "{Spectropolarimetric NLTE inversion code SNAPI}",
      journal = {\aap},
         year = 2018,
        month = sep,
       volume = {617},
          eid = {A24},
        pages = {A24},
          doi = {10.1051/0004-6361/201833382},
archivePrefix = {arXiv},
       eprint = {1806.08134},
 primaryClass = {astro-ph.SR},
       adsurl = {https://ui.adsabs.harvard.edu/abs/2018A&A...617A..24M}
}

@ARTICLE{Michiel_Jaime_2017,
       author = {{de la Cruz Rodr{\'\i}guez}, J. and {van Noort}, M.},
        title = "{Radiative Diagnostics in the Solar Photosphere and Chromosphere}",
      journal = {\ssr},
         year = 2017,
        month = sep,
       volume = {210},
       number = {1-4},
        pages = {109-143},
          doi = {10.1007/s11214-016-0294-8},
archivePrefix = {arXiv},
       eprint = {1609.08324},
 primaryClass = {astro-ph.SR},
       adsurl = {https://ui.adsabs.harvard.edu/abs/2017SSRv..210..109D}
}

@ARTICLE{MURaM2005,
       author = {{V{\"o}gler}, A. and {Shelyag}, S. and {Sch{\"u}ssler}, M. and {Cattaneo}, F. and {Emonet}, T. and {Linde}, T.},
        title = "{Simulations of magneto-convection in the solar photosphere.  Equations, methods, and results of the MURaM code}",
      journal = {\aap},
         year = 2005,
        month = jan,
       volume = {429},
        pages = {335-351},
          doi = {10.1051/0004-6361:20041507},
       adsurl = {https://ui.adsabs.harvard.edu/abs/2005A&A...429..335V}
}

@ARTICLE{NovemberSimon1988,
       author = {{November}, Laurence J. and {Simon}, George W.},
        title = "{Precise Proper-Motion Measurement of Solar Granulation}",
      journal = {\apj},
         year = 1988,
        month = oct,
       volume = {333},
        pages = {427},
          doi = {10.1086/166758},
       adsurl = {https://ui.adsabs.harvard.edu/abs/1988ApJ...333..427N}
}

@ARTICLE{Nordlund_lrsp_2009,
       author = {{Nordlund}, {\r{A}}ke and {Stein}, Robert F. and {Asplund}, Martin},
        title = "{Solar Surface Convection}",
      journal = {Living Rev. Sol. Phys.},
         year = 2009,
        month = dec,
       volume = {6},
       number = {1},
          eid = {2},
        pages = {2},
          doi = {10.12942/lrsp-2009-2},
       adsurl = {https://ui.adsabs.harvard.edu/abs/2009LRSP....6....2N}
}

@ARTICLE{Nordlund_1982_simulations,
       author = {{Nordlund}, A.},
        title = "{Numerical simulations of the solar granulation. I. Basic equations and methods.}",
      journal = {\aap},
         year = 1982,
        month = mar,
       volume = {107},
        pages = {1-10},
       adsurl = {https://ui.adsabs.harvard.edu/abs/1982A&A...107....1N}
}

@ARTICLE{Iglesias2019review,
       author = {{Iglesias}, Francisco A. and {Feller}, Alex},
        title = "{Instrumentation for solar spectropolarimetry: state of the art and prospects}",
      journal = {Opt. Eng.},
         year = 2019,
        month = aug,
       volume = {58},
          eid = {082417},
        pages = {082417},
          doi = {10.1117/1.OE.58.8.082417},
archivePrefix = {arXiv},
       eprint = {1911.01368},
 primaryClass = {astro-ph.IM},
       adsurl = {https://ui.adsabs.harvard.edu/abs/2019OptEn..58h2417I}
}

@ARTICLE{Jarolim2023pinn,
       author = {{Jarolim}, R. and {Thalmann}, J.~K. and {Veronig}, A.~M. and {Podladchikova}, T.},
        title = "{Probing the solar coronal magnetic field with physics-informed neural networks.}",
      journal = {Nat. Astron.},
         year = 2023,
        month = oct,
       volume = {7},
        pages = {1171-1179},
          doi = {10.1038/s41550-023-02030-9},
       adsurl = {https://ui.adsabs.harvard.edu/abs/2023NatAs...7.1171J}
}

@ARTICLE{KorpiLagg2025sunriseIII,
       author = {{Korpi-Lagg}, Andreas and {Gandorfer}, Achim and {Solanki}, Sami K. and {del Toro Iniesta}, Jose Carlos and {Katsukawa}, Yukio and {Bernasconi}, Pietro and {Berkefeld}, Thomas and {Feller}, Alex and {Riethm{\"u}ller}, Tino L. and {{\'A}lvarez-Herrero}, Alberto and {Kubo}, Masahito and {Mart{\'\i}nez Pillet}, Valent{\'\i}n and {Smitha}, H.~N. and {Orozco Su{\'a}rez}, David and {Grauf}, Bianca and {Carpenter}, Michael and {Bell}, Alexander and {{\'A}lvarez-Alonso}, Mar{\'\i}a-Teresa and {{\'A}lvarez Garc{\'\i}a}, Daniel and {Aparicio del Moral}, Beatriz and {Ati{\'e}nzar}, Julia and {Ayoub}, Daniel and {Bail{\'e}n}, Francisco Javier and {Bail{\'o}n Mart{\'\i}nez}, Eduardo and {Balaguer Jim{\'e}nez}, Maria and {Barthol}, Peter and {Bayon Laguna}, Montserrat and {Bellot Rubio}, Luis R. and {Bergmann}, Melani and {Blanco Rodr{\'\i}guez}, Julian and {Bochmann}, Jan and {Borrero}, Juan Manuel and {Campos-Jara}, Antonio and {Castellanos Dur{\'a}n}, Juan Sebasti{\'a}n and {Cebollero}, Mar{\'\i}a and {Conde Rodr{\'\i}guez}, Aitor and {Deutsch}, Werner and {Eaton}, Harry and {Fern{\'a}ndez-Medina}, Ana Belen and {Fernandez-Rico}, German and {Ferreres}, Agustin and {Garc{\'\i}a}, Andr{\'e}s and {Garc{\'\i}a Alarcia}, Ram{\'o}n Mar{\'\i}a and {Garc{\'\i}a Parejo}, Pilar and {Garranzo-Garc{\'\i}a}, Daniel and {Gasent Blesa}, Jos{\'e} Luis and {Gerber}, Karin and {Germerott}, Dietmar and {Gilabert Palmer}, David and {Gizon}, Laurent and {G{\'o}mez S{\'a}nchez-Tirado}, Miguel Angel and {Gonz{\'a}lez-B{\'a}rcena}, David and {Gonzalo Melchor}, Alejandro and {Goodyear}, Sam and {Hara}, Hirohisa and {Harnes}, Edvarda and {Heerlein}, Klaus and {Heidecke}, Frank and {Heinrichs}, Jan and {Hern{\'a}ndez Exp{\'o}sito}, David and {Hirzberger}, Johann and {Hoelken}, Johannes and {Hyun}, Sangwon and {Iglesias}, Francisco A. and {Ishikawa}, Ryohtaroh T. and {Jeon}, Minwoo and {Kawabata}, Yusuke and {Kolleck}, Martin and {Laguna}, Hugo and {Lomas}, Julian and {L{\'o}pez Jim{\'e}nez}, Antonio C. and {Manzano}, Paula and {Matsumoto}, Takuma and {Mayo Turrado}, David and {Meierdierks}, Thimo and {Meining}, Stefan and {Monecke}, Markus and {Morales-Fern{\'a}ndez}, Jos{\'e} Miguel and {Moreno Mantas}, Antonio Jes{\'u}s and {Moreno Vacas}, Alejandro and {M{\"u}ller}, Marc Ferenc and {M{\"u}ller}, Reinhard and {Naito}, Yoshihiro and {Nakai}, Eiji and {N{\'u}{\~n}ez Peral}, Armon{\'\i}a and {Oba}, Takayoshi and {Palo}, Geoffrey and {P{\'e}rez-Grande}, Isabel and {Piqueras Carre{\~n}o}, Javier and {Preis}, Tobias and {Przybylski}, Damien and {Quintero Noda}, Carlos and {Ramanath}, Sandeep and {Ramos M{\'a}s}, Jose Luis and {Raouafi}, Nour and {Rivas-Mart{\'\i}nez}, Mar{\'\i}a-Jes{\'u}s and {Rodr{\'\i}guez Mart{\'\i}nez}, Pedro and {Rodr{\'\i}guez Valido}, Manuel and {Ruiz Cobo}, Basilio and {S{\'a}nchez Rodr{\'\i}guez}, Antonio and {Sanchez Toledo}, Mariano and {S{\'a}nchez G{\'o}mez}, Antonio and {Sanchis Kilders}, Esteban and {Sant}, Kamal and {Santamarina Guerrero}, Pablo and {Schulze}, Erich and {Shimizu}, Toshifumi and {Silva-L{\'o}pez}, Manuel and {Singh}, Kunal and {Siu-Tapia}, Azaymi L. and {Sonner}, Thomas and {Staub}, Jan and {Strecker}, Hanna and {Tobaruela}, Angel and {Torralbo}, Ignacio and {Tritschler}, Alexandra and {Tsuzuki}, Toshihiro and {Uraguchi}, Fumihiro and {Volkmer}, Reiner and {Vourlidas}, Angelos and {Vukadinovi{\'c}}, Du{\v{s}}an and {Werner}, Stephan and {Zerr}, Andreas},
        title = "{SUNRISE III: Overview of Observatory and Instruments}",
      journal = {\solphys},
         year = 2025,
        month = may,
       volume = {300},
       number = {5},
          eid = {75},
        pages = {75},
          doi = {10.1007/s11207-025-02485-1},
archivePrefix = {arXiv},
       eprint = {2502.06483},
 primaryClass = {astro-ph.IM},
       adsurl = {https://ui.adsabs.harvard.edu/abs/2025SoPh..300...75K}
}

@ARTICLE{Leenaarts2025hesp,
       author = {{Leenaarts}, J. and {van Noort}, M. and {de la Cruz Rodr{\'\i}guez}, J. and {Danilovic}, S. and {D{\'\i}az Baso}, C.~J. and {Hillberg}, T. and {S{\"u}tterlin}, P. and {Kiselman}, D. and {Scharmer}, G. and {Solanki}, S.~K.},
        title = "{High flow speeds and transition-region-like temperatures in the solar chromosphere during flux emergence: Evidence from imaging spectropolarimetry in He I 1083 nm and numerical simulations}",
      journal = {\aap},
         year = 2025,
        month = apr,
       volume = {696},
          eid = {A3},
        pages = {A3},
          doi = {10.1051/0004-6361/202453355},
archivePrefix = {arXiv},
       eprint = {2501.10246},
 primaryClass = {astro-ph.SR},
       adsurl = {https://ui.adsabs.harvard.edu/abs/2025A&A...696A...3L}
}

@ARTICLE{vanNoort2022MiHI,
       author = {{van Noort}, M. and {Bischoff}, J. and {Kramer}, A. and {Solanki}, S.~K. and {Kiselman}, D.},
        title = "{A prototype of a microlensed hyperspectral imager for solar observations}",
      journal = {\aap},
         year = 2022,
        month = dec,
       volume = {668},
          eid = {A149},
        pages = {A149},
          doi = {10.1051/0004-6361/202243464},
       adsurl = {https://ui.adsabs.harvard.edu/abs/2022A&A...668A.149V}
}

@ARTICLE{Schou2012,
       author = {{Schou}, J. and {Scherrer}, P.~H. and {Bush}, R.~I. and {Wachter}, R. and {Couvidat}, S. and {Rabello-Soares}, M.~C. and {Bogart}, R.~S. and {Hoeksema}, J.~T. and {Liu}, Y. and {Duvall}, T.~L. and {Akin}, D.~J. and {Allard}, B.~A. and {Miles}, J.~W. and {Rairden}, R. and {Shine}, R.~A. and {Tarbell}, T.~D. and {Title}, A.~M. and {Wolfson}, C.~J. and {Elmore}, D.~F. and {Norton}, A.~A. and {Tomczyk}, S.},
        title = "{Design and Ground Calibration of the Helioseismic and Magnetic Imager (HMI) Instrument on the Solar Dynamics Observatory (SDO)}",
      journal = {\solphys},
         year = 2012,
        month = jan,
       volume = {275},
       number = {1-2},
        pages = {229-259},
          doi = {10.1007/s11207-011-9842-2},
       adsurl = {https://ui.adsabs.harvard.edu/abs/2012SoPh..275..229S}
}

@ARTICLE{Smitha_nlteI,
       author = {{Smitha}, H.~N. and {Holzreuter}, R. and {van Noort}, M. and {Solanki}, S.~K.},
        title = "{The influence of NLTE effects in Fe I lines on an inverted atmosphere. I. 6301 {\r{A}} and 6302 {\r{A}} lines formed in 1D NLTE}",
      journal = {\aap},
         year = 2020,
        month = jan,
       volume = {633},
          eid = {A157},
        pages = {A157},
          doi = {10.1051/0004-6361/201937041},
archivePrefix = {arXiv},
       eprint = {1912.07007},
 primaryClass = {astro-ph.SR},
       adsurl = {https://ui.adsabs.harvard.edu/abs/2020A&A...633A.157S}
}

@ARTICLE{Stein2012,
       author = {{Stein}, Robert F.},
        title = "{Solar Surface Magneto-Convection}",
      journal = {Living Rev. Sol. Phys.},
         year = 2012,
        month = dec,
       volume = {9},
       number = {1},
          eid = {4},
        pages = {4},
          doi = {10.12942/lrsp-2012-4},
       adsurl = {https://ui.adsabs.harvard.edu/abs/2012LRSP....9....4S}
}

@ARTICLE{Rempel2014,
       author = {{Rempel}, M.},
        title = "{Numerical Simulations of Quiet Sun Magnetism: On the Contribution from a Small-scale Dynamo}",
      journal = {\apj},
         year = 2014,
        month = jul,
       volume = {789},
       number = {2},
          eid = {132},
        pages = {132},
          doi = {10.1088/0004-637X/789/2/132},
archivePrefix = {arXiv},
       eprint = {1405.6814},
 primaryClass = {astro-ph.SR},
       adsurl = {https://ui.adsabs.harvard.edu/abs/2014ApJ...789..132R}
}

@ARTICLE{ERempel2022tracking,
       author = {{Rempel}, Erico L. and {Chertovskih}, Roman and {Davletshina}, Kamilla R. and {Silva}, Suzana S.~A. and {Welsch}, Brian T. and {Chian}, Abraham C. -L.},
        title = "{Reconstruction of Photospheric Velocity Fields from Highly Corrupted Data}",
      journal = {\apj},
         year = 2022,
        month = jul,
       volume = {933},
       number = {1},
          eid = {2},
        pages = {2},
          doi = {10.3847/1538-4357/ac6fe4},
archivePrefix = {arXiv},
       eprint = {2205.09846},
 primaryClass = {astro-ph.SR},
       adsurl = {https://ui.adsabs.harvard.edu/abs/2022ApJ...933....2R}
}

@ARTICLE{Rieutord2001,
       author = {{Rieutord}, M. and {Roudier}, T. and {Ludwig}, H. -G. and {Nordlund}, {\r{A}}. and {Stein}, R.},
        title = "{Are granules good tracers of solar surface velocity fields?}",
      journal = {\aap},
         year = 2001,
        month = oct,
       volume = {377},
        pages = {L14-L17},
          doi = {10.1051/0004-6361:20011160},
archivePrefix = {arXiv},
       eprint = {astro-ph/0108284},
 primaryClass = {astro-ph},
       adsurl = {https://ui.adsabs.harvard.edu/abs/2001A&A...377L..14R}
}

@ARTICLE{Schuck_2006_DAVE,
       author = {{Schuck}, P.~W.},
        title = "{Tracking Magnetic Footpoints with the Magnetic Induction Equation}",
      journal = {\apj},
         year = 2006,
        month = aug,
       volume = {646},
       number = {2},
        pages = {1358-1391},
          doi = {10.1086/505015},
       adsurl = {https://ui.adsabs.harvard.edu/abs/2006ApJ...646.1358S}
}

@ARTICLE{Sunrise:2010,
       author = {{Solanki}, S.~K. and {Barthol}, P. and {Danilovic}, S. and {Feller}, A. and {Gandorfer}, A. and {Hirzberger}, J. and {Riethm{\"u}ller}, T.~L. and {Sch{\"u}ssler}, M. and {Bonet}, J.~A. and {Mart{\'\i}nez Pillet}, V. and {del Toro Iniesta}, J.~C. and {Domingo}, V. and {Palacios}, J. and {Kn{\"o}lker}, M. and {Bello Gonz{\'a}lez}, N. and {Berkefeld}, T. and {Franz}, M. and {Schmidt}, W. and {Title}, A.~M.},
        title = "{SUNRISE: Instrument, Mission, Data, and First Results}",
      journal = {\apjl},
         year = 2010,
        month = nov,
       volume = {723},
       number = {2},
        pages = {L127-L133},
          doi = {10.1088/2041-8205/723/2/L127},
archivePrefix = {arXiv},
       eprint = {1008.3460},
 primaryClass = {astro-ph.SR},
       adsurl = {https://ui.adsabs.harvard.edu/abs/2010ApJ...723L.127S}
}

@ARTICLE{SUNRISE,
       author = {{Barthol}, P. and {Gandorfer}, A. and {Solanki}, S.~K. and {Sch{\"u}ssler}, M. and {Chares}, B. and {Curdt}, W. and {Deutsch}, W. and {Feller}, A. and {Germerott}, D. and {Grauf}, B. and {Heerlein}, K. and {Hirzberger}, J. and {Kolleck}, M. and {Meller}, R. and {M{\"u}ller}, R. and {Riethm{\"u}ller}, T.~L. and {Tomasch}, G. and {Kn{\"o}lker}, M. and {Lites}, B.~W. and {Card}, G. and {Elmore}, D. and {Fox}, J. and {Lecinski}, A. and {Nelson}, P. and {Summers}, R. and {Watt}, A. and {Mart{\'\i}nez Pillet}, V. and {Bonet}, J.~A. and {Schmidt}, W. and {Berkefeld}, T. and {Title}, A.~M. and {Domingo}, V. and {Gasent Blesa}, J.~L. and {del Toro Iniesta}, J.~C. and {L{\'o}pez Jim{\'e}nez}, A. and {{\'A}lvarez-Herrero}, A. and {Sabau-Graziati}, L. and {Widani}, C. and {Haberler}, P. and {H{\"a}rtel}, K. and {Kampf}, D. and {Levin}, T. and {P{\'e}rez Grande}, I. and {Sanz-Andr{\'e}s}, A. and {Schmidt}, E.},
        title = "{The Sunrise Mission}",
      journal = {\solphys},
         year = 2011,
        month = jan,
       volume = {268},
       number = {1},
        pages = {1-34},
          doi = {10.1007/s11207-010-9662-9},
archivePrefix = {arXiv},
       eprint = {1009.2689},
 primaryClass = {astro-ph.IM},
       adsurl = {https://ui.adsabs.harvard.edu/abs/2011SoPh..268....1B}
}

@ARTICLE{SiuTapia2025MgIb1,
       author = {{Siu-Tapia}, A.~L. and {Bellot Rubio}, L.~R. and {Orozco Su{\'a}rez}, D.},
        title = "{Diagnosing the solar atmosphere through the Mg I b$_{2}$ 5173 {\r{A}} line: I. Nonlocal thermodynamic equilibrium inversions versus traditional inferences}",
      journal = {\aap},
         year = 2025,
        month = apr,
       volume = {696},
          eid = {A105},
        pages = {A105},
          doi = {10.1051/0004-6361/202453230},
archivePrefix = {arXiv},
       eprint = {2503.12498},
 primaryClass = {astro-ph.SR},
       adsurl = {https://ui.adsabs.harvard.edu/abs/2025A&A...696A.105S}
}

@ARTICLE{SiuTapia2025MgIb2,
       author = {{Siu-Tapia}, A.~L. and {Bellot Rubio}, L.~R. and {Orozco Su{\'a}rez}, D. and {Gafeira}, R.},
        title = "{Diagnosing the solar atmosphere through the Mg I b$_{2}$ 5173 {\r{A}} line: II. Morphological classification of the intensity and circular polarization profiles}",
      journal = {\aap},
         year = 2025,
        month = apr,
       volume = {696},
          eid = {A106},
        pages = {A106},
          doi = {10.1051/0004-6361/202453232},
archivePrefix = {arXiv},
       eprint = {2503.12501},
 primaryClass = {astro-ph.SR},
       adsurl = {https://ui.adsabs.harvard.edu/abs/2025A&A...696A.106S}
}

@ARTICLE{swirl2019,
       author = {{Liu}, Jiajia and {Nelson}, Chris J. and {Snow}, Ben and {Wang}, Yuming and {Erd{\'e}lyi}, Robert},
        title = "{Evidence of ubiquitous Alfv{\'e}n pulses transporting energy from the photosphere to the upper chromosphere}",
      journal = {Nat. Commun.},
         year = 2019,
        month = aug,
       volume = {10},
          eid = {3504},
        pages = {3504},
          doi = {10.1038/s41467-019-11495-0},
       adsurl = {https://ui.adsabs.harvard.edu/abs/2019NatCo..10.3504L}
}

@ARTICLE{Skirvin_2024ApJ_MHDwaves,
       author = {{Skirvin}, Samuel J. and {Fedun}, Viktor and {Goossens}, Marcel and {Silva}, Suzana S.~A. and {Verth}, Gary},
        title = "{Poynting Flux of MHD Modes in Magnetic Solar Vortex Tubes}",
      journal = {\apj},
         year = 2024,
        month = nov,
       volume = {975},
       number = {2},
          eid = {176},
        pages = {176},
          doi = {10.3847/1538-4357/ad7de1},
archivePrefix = {arXiv},
       eprint = {2410.05204},
 primaryClass = {astro-ph.SR},
       adsurl = {https://ui.adsabs.harvard.edu/abs/2024ApJ...975..176S}
}

@ARTICLE{Tilipman2023,
       author = {{Tilipman}, Dennis and {Kazachenko}, Maria and {Tremblay}, Benoit and {Mili{\'c}}, Ivan and {Mart{\'\i}nez Pillet}, Valentin and {Rempel}, Matthias},
        title = "{Quantifying Poynting Flux in the Quiet Sun Photosphere}",
      journal = {\apj},
         year = 2023,
        month = oct,
       volume = {956},
       number = {2},
          eid = {83},
        pages = {83},
          doi = {10.3847/1538-4357/ace621},
archivePrefix = {arXiv},
       eprint = {2307.02445},
 primaryClass = {astro-ph.SR},
       adsurl = {https://ui.adsabs.harvard.edu/abs/2023ApJ...956...83T}
}

@INPROCEEDINGS{Tremblay2023,
       author = {{Tremblay}, Benoit and {Jarolim}, Robert and {Rempel}, Matthias and {Molnar}, Momchil and {Thalmann}, Julia and {Veronig}, Astrid and {Podladchikova}, Tatiana},
        title = "{Physics-informed neural networks to model magnetic fields and velocity fields}",
    booktitle = {54th Meeting of the Solar Physics Division},
         year = 2023,
       series = {AAS/Solar Physics Division Meeting},
       volume = {55},
        month = oct,
          eid = {202.01},
        pages = {202.01},
       adsurl = {https://ui.adsabs.harvard.edu/abs/2023SPD....5420201T}
}

@ARTICLE{TuMag,
       author = {{del Toro Iniesta}, J.~C. and {Orozco Su{\'a}rez}, D. and {{\'A}lvarez-Herrero}, A. and {Sanchis Kilders}, E. and {P{\'e}rez-Grande}, I. and {Ruiz Cobo}, B. and {Bellot Rubio}, L.~R. and {Balaguer Jim{\'e}nez}, M. and {L{\'o}pez Jim{\'e}nez}, A.~C. and {{\'A}lvarez Garc{\'\i}a}, D. and {Ramos M{\'a}s}, J.~L. and {Cobos Carrascosa}, J.~P. and {Labrousse}, P. and {Moreno Mantas}, A.~J. and {Morales-Fern{\'a}ndez}, J.~M. and {Aparicio del Moral}, B. and {S{\'a}nchez G{\'o}mez}, A. and {Bail{\'o}n Mart{\'\i}nez}, E. and {Bail{\'e}n}, F.~J. and {Strecker}, H. and {Siu-Tapia}, A.~L. and {Santamarina Guerrero}, P. and {Moreno Vacas}, A. and {Ati{\'e}nzar Garc{\'\i}a}, J. and {Dorantes Monteagudo}, A.~J. and {Bustamante}, I. and {Tobaruela}, A. and {Fern{\'a}ndez-Medina}, A. and {N{\'u}{\~n}ez Peral}, A. and {Cebollero}, M. and {Garranzo-Garc{\'\i}a}, D. and {Garc{\'\i}a Parejo}, P. and {Gonzalo Melchor}, A. and {S{\'a}nchez Rodr{\'\i}guez}, A. and {Campos-Jara}, A. and {Laguna}, H. and {Silva-L{\'o}pez}, M. and {Blanco Rodr{\'\i}guez}, J. and {Gasent Blesa}, J.~L. and {Rodr{\'\i}guez Mart{\'\i}nez}, P. and {Ferreres}, A. and {Gilabert Palmer}, D. and {Torralbo}, I. and {Piqueras}, J. and {Gonz{\'a}lez-B{\'a}rcena}, D. and {Fern{\'a}ndez}, A.~J. and {Hern{\'a}ndez Exp{\'o}sito}, D. and {P{\'a}ez Ma{\~n}{\'a}}, E. and {Magdaleno Castell{\'o}}, E. and {Rodr{\'\i}guez Valido}, M. and {Korpi-Lagg}, Andreas and {Gandorfer}, Achim and {Solanki}, Sami K. and {Berkefeld}, Thomas and {Bernasconi}, Pietro and {Feller}, Alex and {Katsukawa}, Yukio and {Riethm{\"u}ller}, Tino L. and {Smitha}, H.~N. and {Kubo}, Masahito and {Mart{\'\i}nez Pillet}, Valent{\'\i}n and {Grauf}, Bianca and {Bell}, Alexander and {Carpenter}, Michael},
        title = "{TuMag: the tunable magnetograph for the Sunrise III mission}",
      journal = {arXiv e-prints},
         year = 2025,
        month = feb,
          eid = {arXiv:2502.08268},
        pages = {arXiv:2502.08268},
          doi = {10.48550/arXiv.2502.08268},
archivePrefix = {arXiv},
       eprint = {2502.08268},
 primaryClass = {astro-ph.IM},
       adsurl = {https://ui.adsabs.harvard.edu/abs/2025arXiv250208268D}
}

@ARTICLE{Verma2013,
       author = {{Verma}, M. and {Steffen}, M. and {Denker}, C.},
        title = "{Evaluating local correlation tracking using CO5BOLD simulations of solar granulation}",
      journal = {\aap},
         year = 2013,
        month = jul,
       volume = {555},
          eid = {A136},
        pages = {A136},
          doi = {10.1051/0004-6361/201321628},
archivePrefix = {arXiv},
       eprint = {1305.6033},
 primaryClass = {astro-ph.SR},
       adsurl = {https://ui.adsabs.harvard.edu/abs/2013A&A...555A.136V}
}

@ARTICLE{Vukadinovic2022mg,
       author = {{Vukadinovi{\'c}}, D. and {Mili{\'c}}, I. and {Atanackovi{\'c}}, O.},
        title = "{Investigating magnetic field inference from the spectral region around the Mg I b$_{2}$ line using the weak-field approximation}",
      journal = {\aap},
         year = 2022,
        month = aug,
       volume = {664},
          eid = {A182},
        pages = {A182},
          doi = {10.1051/0004-6361/202142015},
archivePrefix = {arXiv},
       eprint = {2205.04236},
 primaryClass = {astro-ph.SR},
       adsurl = {https://ui.adsabs.harvard.edu/abs/2022A&A...664A.182V}
}

@INPROCEEDINGS{vtf,
       author = {{Schmidt}, Wolfgang and {Schubert}, Matthias and {Ellwarth}, Monika and {Baumgartner}, J{\"o}rg and {Bell}, Alexander and {Fischer}, Andreas and {Halbgewachs}, Clemens and {Heidecke}, Frank and {Kentischer}, Thomas and {von der L{\"u}he}, Oskar and {Scheiffelen}, Thomas and {Sigwarth}, Michael},
        title = "{End-to-end simulations of the visible tunable filter for the Daniel K. Inouye Solar Telescope}",
    booktitle = {Ground-based and Airborne Instrumentation for Astronomy VI},
         year = 2016,
       editor = {{Evans}, Christopher J. and {Simard}, Luc and {Takami}, Hideki},
       series = {Society of Photo-Optical Instrumentation Engineers (SPIE) Conference Series},
       volume = {9908},
        month = aug,
          eid = {99084N},
        pages = {99084N},
          doi = {10.1117/12.2232518},
archivePrefix = {arXiv},
       eprint = {1607.06767},
 primaryClass = {astro-ph.IM},
       adsurl = {https://ui.adsabs.harvard.edu/abs/2016SPIE.9908E..4NS}
}

@article{Welsch2004,
    doi = {10.1086/421767},
    url = {https://dx.doi.org/10.1086/421767},
    year = {2004},
    month = {aug},
    publisher = {},
    volume = {610},
    number = {2},
    pages = {1148},
    author = {B. T. Welsch and G. H. Fisher and W. P. Abbett and S. Regnier},
    title = {ILCT: Recovering Photospheric Velocities from Magnetograms by Combining the Induction Equation with Local Correlation Tracking},
    journal = {\apj}
}

@ARTICLE{Welsch2007,
       author = {{Welsch}, B.~T. and {Abbett}, W.~P. and {De Rosa}, M.~L. and {Fisher}, G.~H. and {Georgoulis}, M.~K. and {Kusano}, K. and {Longcope}, D.~W. and {Ravindra}, B. and {Schuck}, P.~W.},
        title = "{Tests and Comparisons of Velocity-Inversion Techniques}",
      journal = {\apj},
         year = 2007,
        month = dec,
       volume = {670},
       number = {2},
        pages = {1434-1452},
          doi = {10.1086/522422},
       adsurl = {https://ui.adsabs.harvard.edu/abs/2007ApJ...670.1434W}
}

@INPROCEEDINGS{Welsch2008,
       author = {{Fisher}, G.~H. and {Welsch}, B.~T.},
        title = "{FLCT: A Fast, Efficient Method for Performing Local Correlation Tracking}",
    booktitle = {Subsurface and Atmospheric Influences on Solar Activity},
         year = 2008,
       editor = {{Howe}, R. and {Komm}, R.~W. and {Balasubramaniam}, K.~S. and {Petrie}, G.~J.~D.},
       series = {Astronomical Society of the Pacific Conference Series},
       volume = {383},
        month = jan,
        pages = {373},
          doi = {10.48550/arXiv.0712.4289},
archivePrefix = {arXiv},
       eprint = {0712.4289},
 primaryClass = {astro-ph},
       adsurl = {https://ui.adsabs.harvard.edu/abs/2008ASPC..383..373F}
}

@article{Welsch2012,
doi = {10.1088/0004-637X/747/2/130},
url = {https://dx.doi.org/10.1088/0004-637X/747/2/130},
year = {2012},
month = {feb},
publisher = {The American Astronomical Society},
volume = {747},
number = {2},
pages = {130},
author = {B. T. Welsch and K. Kusano and T. T. Yamamoto and K. Muglach},
title = {DECORRELATION TIMES OF PHOTOSPHERIC FIELDS AND FLOWS},
journal = {\apj}
}

@ARTICLE{Welsch_2015_plage_poyinting,
       author = {{Welsch}, Brian T.},
        title = "{The photospheric Poynting flux and coronal heating}",
      journal = {\pasj},
         year = 2015,
        month = apr,
       volume = {67},
       number = {2},
          eid = {18},
        pages = {18},
          doi = {10.1093/pasj/psu151},
archivePrefix = {arXiv},
       eprint = {1402.4794},
 primaryClass = {astro-ph.SR},
       adsurl = {https://ui.adsabs.harvard.edu/abs/2015PASJ...67...18W}
}

@ARTICLE{danilovic2016,
       author = {{Danilovic}, S. and {Rempel}, M. and {van Noort}, M. and {Cameron}, R.},
        title = "{Observed and simulated power spectra of kinetic and magnetic energy retrieved with 2D inversions}",
      journal = {\aap},
         year = 2016,
        month = oct,
       volume = {594},
          eid = {A103},
        pages = {A103},
          doi = {10.1051/0004-6361/201527917},
archivePrefix = {arXiv},
       eprint = {1607.06242},
 primaryClass = {astro-ph.SR},
       adsurl = {https://ui.adsabs.harvard.edu/abs/2016A&A...594A.103D}
}

%-----------------------------------------------------------

\end{document}